\documentclass{vldb}
\usepackage{subfig}
\usepackage{graphicx}
\usepackage{balance}  
\usepackage[linesnumbered,ruled,vlined]{algorithm2e}
\usepackage{multirow}
\usepackage{changepage}
\DeclareMathOperator*{\argmax}{arg\,max}

\usepackage{amsthm}

\vldbTitle{Efficient Exact Algorithms for Maximum Balanced Biclique Search in Bipartite Graphs}
\vldbAuthors{Lu Chen, Chengfei Liu, Rui Zhou, Jiajie Xu, Jianxin Li}
\vldbDOI{https://doi.org/10.14778/xxxxxxx.xxxxxxx}
\vldbVolume{xx}
\vldbNumber{xxx}
\vldbYear{xxxx}

\begin{document}


\title{Efficient Exact Algorithms for Maximum Balanced Biclique Search in Bipartite Graphs}



%
%
%
%

\numberofauthors{1} 

\author{
    \alignauthor
Lu Chen\textsuperscript{$\dagger$}, Chengfei Liu\textsuperscript{$\dagger$}, Rui Zhou\textsuperscript{$\dagger$}, Jiajie Xu\textsuperscript{$\S$}, Jianxin Li\textsuperscript{$\P$} \\
\textsuperscript{$\dagger$}\affaddr{Swinburne University of Technology},
\textsuperscript{$\S$}\affaddr{Soochow University},  \textsuperscript{$\P$}\affaddr{Deakin University}\\
\email{\textsuperscript{$\dagger$}\{luchen, cliu, rzhou\}@swin.edu.au \textsuperscript{$\S$}xujj@suda.edu.cn \textsuperscript{$\P$}jianxin.li@deakin.edu.au
}}

%
%

\date{30 July 1999}

\maketitle

\begin{abstract}
Given a bipartite graph, the maximum balanced biclique (\textsf{MBB}) problem, discovering a mutually connected while equal-sized disjoint sets with the maximum cardinality, plays a significant role for mining the bipartite graph and has numerous applications.
Despite the NP-hardness of the \textsf{MBB} problem, in this paper, we show that an exact \textsf{MBB} can be discovered extremely fast in bipartite graphs for real applications. We propose two exact algorithms dedicated for dense and sparse bipartite graphs respectively.
For dense bipartite graphs, an $\mathcal{O}^{*}( 1.3803^{n})$ algorithm is proposed.
This algorithm in fact can find an \textsf{MBB} in near polynomial time for dense bipartite graphs that are common for applications such as VLSI design.
This is because, using our proposed novel techniques, the search can fast converge to sufficiently dense bipartite graphs which we prove to be polynomially solvable.
For large sparse bipartite graphs typical for applications such as biological data analysis, an $\mathcal{O}^{*}( 1.3803^{\ddot{\delta}})$ algorithm is proposed, where $\ddot{\delta}$ is only a few hundreds for large sparse bipartite graphs with millions of vertices.
The indispensible optimizations that lead to this time complexity are: we transform a large sparse bipartite graph into a limited number of dense subgraphs with size up to $\ddot{\delta}$ and then apply our proposed algorithm for dense bipartite graphs on each of the subgraphs.
To further speed up this algorithm, tighter upper bounds, faster heuristics and effective reductions are proposed, allowing an \textsf{MBB} to be discovered within a few seconds for bipartite graphs with millions of vertices.
Extensive experiments are conducted on synthetic and real large bipartite graphs to demonstrate the efficiency and effectiveness of our proposed algorithms and techniques.

\end{abstract}

\newtheorem*{rproblem}{Problem}
\newtheorem{definition}{Definition}
\newtheorem{lemma}{Lemma}
\newtheorem{obs}{Observation} 
\section{Introduction}

\noindent\textbf{Maximum balanced biclique problem}. Given a bipartite graph $G=(L,R,E)$, a biclique $(A\subseteq L ,B\subseteq R)$ is a subgraph of $G$ such that $\forall$ $2$-tuple $(u,v)\in A\times B$, $(u,v)\in E$.
When $|A|=|B|$, $(A,B)$ is a balanced biclique.
One of the fundamental but significant biclique problems is the maximum balanced biclique (\textsf{MBB}) problem, i.e., given  a bipartite graph $G$, finding a balanced biclique with the maximum number of vertices.

\noindent\textbf{Significance}. 
The \textsf{MBB} problem is significant across various disciplines.
It has extensive real applications for very-large-scale integration (VLSI) including programmable logic array folding~\cite{ravi1988complexity}, defect tolerance chips designing~\cite{al2007defect,tahoori2006application}, etc.
Furthermore, it plays principal roles for analyzing biological data since an \textsf{MBB} is an important instance of bicluster~\cite{cheng2000biclustering,yang2005improved}.
Recently, it also draws significant attention for discovering interactions between proteins~\cite{bustamam2020application,mukhopadhyay2014incorporating,dey2019graph,kaloka2019pols}.

Given the significance of the \textsf{MBB} problem, it has been studied extensively.
Since the \textsf{MBB} problem has been proven to be NP-hard \cite{garey1979computers} and NP-hard to approximate within $n^{1-\varepsilon}$ factor for every $\varepsilon>0$  \cite{manurangsi2018inapproximability}, most existing algorithms~\cite{ZHOU201986,WU2015693,LI2020104922,wang2018new} for finding an \textsf{MBB} are heuristic algorithms while very few works were dedicated for finding an exact \textsf{MBB}, other than the work \cite{ZHOU2018834}.

\noindent\textbf{Our approach}. In this paper, we focus on finding an exact \textsf{MBB}.
Surprisingly, we show that an exact \textsf{MBB} can be discovered extremely fast, by making the benefit of the characteristics of the bipartite graphs of real applications, despite the NP-hardness of the \textsf{MBB} problem.
We first make a breakthrough in solving the \textsf{MBB} problem for dense bipartite graphs and then devise an efficient algorithm for large sparse bipartite graphs by taking the advantages of our algorithm for dense bipartite graphs.
The prominent motivations and intuitions of our proposed algorithms are introduced as follows.

\noindent\underline{\textit{Novel algorithm for dense bipartite graphs}}.
We observe that bipartite graphs are quite dense in applications such as VLSI design. Finding exact results for these applications would significantly improve the robustness of the designed circuit.
However, the existing exact \textsf{MBB} algorithm \cite{ZHOU2018834} cannot find the exact result within a few hours for dense bipartite graphs with just hundreds of vertices. 
This motivates us to devise novel techniques for dealing with dense bipartite graphs, leading to a novel algorithm denoted as \textsf{denseMBB}.
\textsf{denseMBB} has a time complexity of $\mathcal{O}^{*}( 1.3803^{n})$ where $n$ is the number of vertices in $G$.
To the best of our knowledge, \textsf{denseMBB} is the first \textsf{MBB} algorithm for speeding up \textsf{MBB} search in dense bipartite graphs with explicit time complexity.
The intuitions of \textsf{denseMBB} are below.
We propose a polynomial algorithm for finding an exact \textsf{MBB} when a bipartite graph is sufficiently dense.
Then, triviality last branching strategy is given to avoid enumerating on subgraphs where our proposed polynomial algorithm can apply.
In fact, \textsf{denseMBB} can find an \textsf{MBB} in near polynomial time for a bipartite graph $G=(L,R,E)$ where $|E|$ is around $80\%$ to $90\%$ of $|L|\times |R|$, which is quite typical for defect tolerance chips designing~\cite{al2007defect,tahoori2006application}.
This is because, when bipartite graphs are dense, our proposed novel techniques make the search converge to polynomially solvable subgraphs in near constant steps.

\noindent\underline{\textit{Novel algorithm for large sparse bipartite graphs}}.
Given the promising time complexity of \textsf{denseMBB}, it is natural to ask whether \textsf{denseMBB} can be applied to large sparse bipartite graphs that are typical for applications such as analyzing biological data.
Applying \textsf{denseMBB} to large sparse bipartite graphs directly is inefficient in practice given the fact that the number of vertices could be extremely large and optimizations for dense bipartite graph cannot significantly reduce $\mathcal{O}^{*}( 1.3803^{n})$ when bipartite graphs are sparse.
We propose a novel algorithm \textsf{sparseMBB} for dealing with large sparse bipartite graphs with time complexity of $\mathcal{O}^{*}( 1.3803^{\ddot{\delta}})$, where $\ddot{\delta}$ is a novel bipartite sparsity parameter proposed by us and is only a few hundreds for large sparse bipartite graphs having millions vertices.
The intuitions of \textsf{sparseMBB} are as follows.
Using our proposed techniques, \textsf{sparseMBB} transforms a large bipartite graph into a limited number of small but dense subgraphs with size up to $\ddot{\delta}$.
After that, our proposed \textsf{denseMBB} is applied to each small but dense subgraphs, which makes $\mathcal{O}^{*}( 1.3803^{\ddot{\delta}})$ near polynomial in practice.
Apart from theoretically promising, \textsf{sparseMBB} is very fast practically. In fact, \textsf{sparseMBB} can find an \textsf{MBB} within a few seconds for million-vertex bipartite graphs. 

\begin{figure}
\centering
    \subfloat[Dense]{\includegraphics[height=2.1cm]{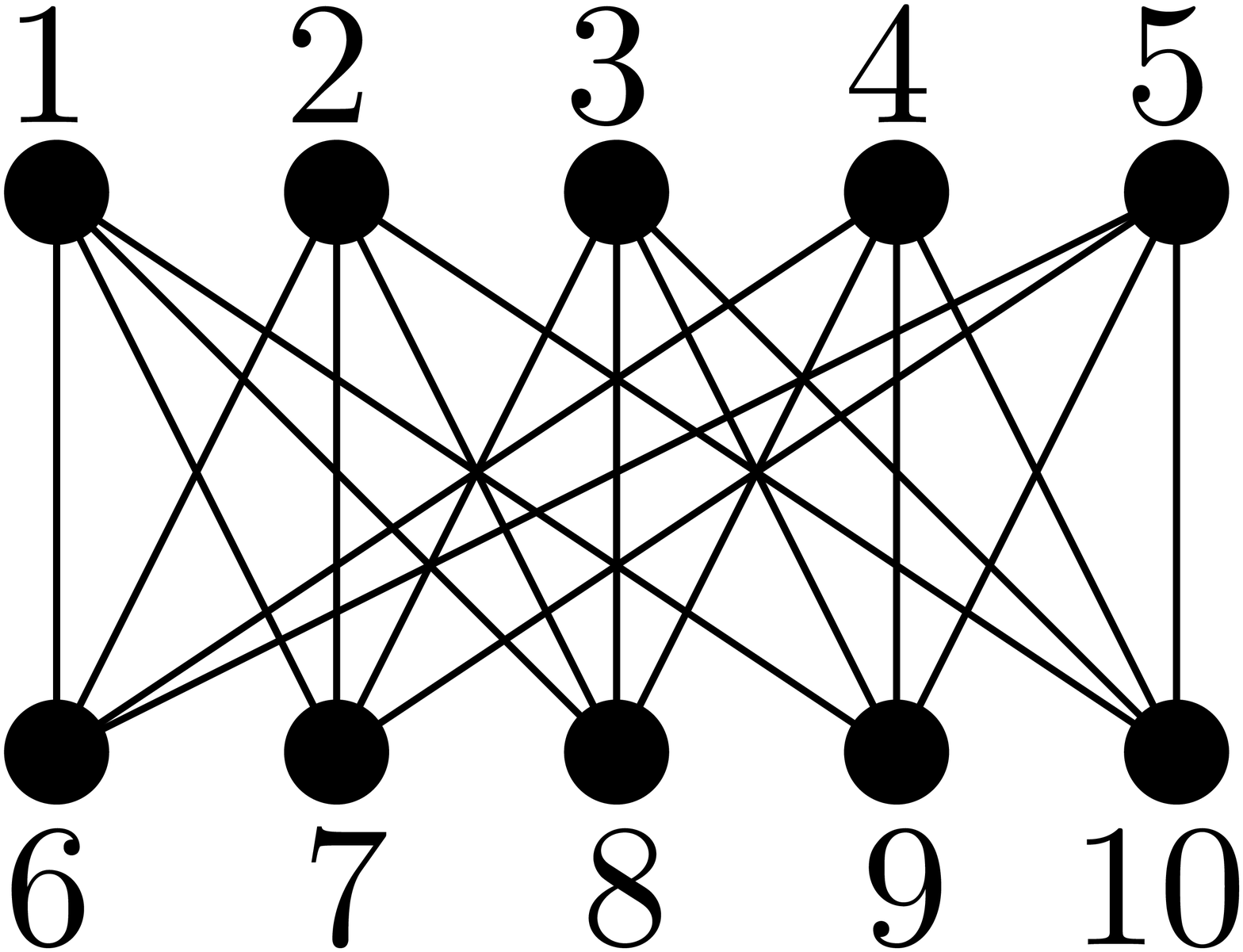}}
	\hspace{10pt}
	\subfloat[Sparse]{\includegraphics[height=2.1cm]{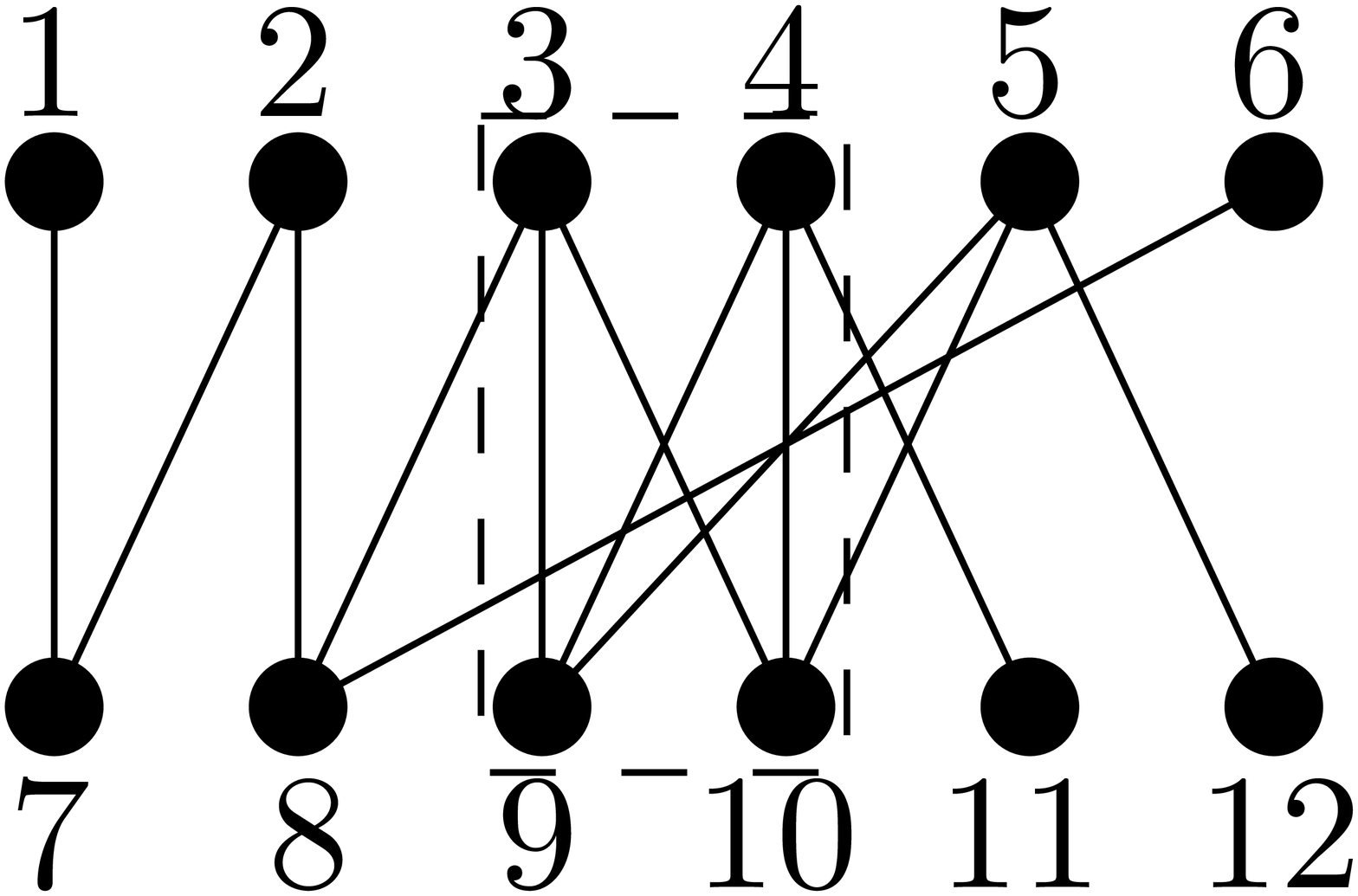}}
	\caption{Dense and sparse bipartite graphs}\label{fig:example}
\vspace{-15pt}
\end{figure}

We highlight our principal contributions below.
\begin{itemize}
  \item Theoretical contributions:
  \vspace{-7pt}
    \begin{itemize}
      \item Novel bipartite sparsity measurement: bipartite degeneracy is proposed for measuring the bipartite sparsity of a bipartite graph, denoted as $\ddot{\delta}$.
      \item Algorithms with better time complexity: our proposed algorithms find an exact result with time complexity of $\mathcal{O}^{*}$$(1.3803^{n})$ for dense bipartite graphs and $\mathcal{O}^{*}( 1.3803^{\ddot{\delta}})$ for large sparse bipartite graphs.
    \end{itemize}
  \item Practically fast algorithms: we conduct extensive experiments on synthetic and real datasets. Our algorithms are up to several order faster than the state-of-the-art algorithm and a number of non-trivial baselines.
\end{itemize}

\noindent\textbf{Roadmap}. The remaining of the paper is organized as follows.
Section \ref{sec:pd} formally defines the \textsf{MBB} problem.
Section \ref{sec:sota} discusses the state-of-the-art algorithm.
Section \ref{sec:dense} introduces our novel algorithm, \textsf{denseMBB} for dense bipartite graphs.
Section \ref{sec:sparse} introduces our novel algorithm, \textsf{sparseMBB} for large sparse bipartite graphs.
Section \ref{sec:exp} experimentally evaluates the efficiency of our proposed algorithms.
Section \ref{sec:rw} discusses related works and Section \ref{sec:con} concludes the paper.

\section{Preliminary and Problem formulation}\label{sec:pd}

Frequently used notations are summarized in Table~\ref{tab:notations}. 

\begin{table}
\centering
\caption{Notations}\label{tab:notations}
\vspace{-10pt}
\begin{adjustwidth}{-0.5cm}{}
\small
\begin{tabular}{r|l}
	\hline
	Notation & Explanation \\
	\hline
	$G=(L,R,E)$ & a bipartite graph\\
	\hline
    $V(G)$ & $L\cup R$\\
	\hline
	$(A,B)$ & a (partial) biclique with $A\subseteq L$ and $B\subseteq B$\\
	\hline
	$(A',B')$, $(A^{*},B^{*})$ & a biclique/bipartite graph\\
    \hline
    $a$, $b$ & $|A|$ and $|B|$ of a biclique \\
    \hline
    $(a,b)$ & parameters for size-constraint biclique problem\\
	\hline
	$H$, $G'$, $G''$, $G'''$  & a subgraph of $G$\\
	\hline
	$\mathcal{H}$ &  a set of subgraphs of $G$\\
	\hline
	$L(H)$, $R(H)$ & disjoint vertex sets of $H$\\
	\hline
	$u$, $v$ 	& a vertex of $G$ \\
	\hline
	$N(u,H)$	& neighbours of $u$ in $H$\\
	\hline
	$N_{2}(u,H)$ & vertices 2-hop from $u$ in $H$\\
	\hline
	$N_{\le 2}(u,H)$ & $N(u,H)$ $\cup$ $N_{2}(u,H)$\\
	\hline
	$\delta(G)$ 	& degeneracy of $G$\\
	\hline
	$\ddot{\delta}(G)$ & bipartite degeneracy of $G$\\
	\hline
	$d_{max}$& maximum degree of $G$ \\
	\hline	
\end{tabular}
\end{adjustwidth}
\end{table}

\noindent\textbf{Bipartite graph}. A bipartite graph is a graph in which vertices can be partitioned into two sets $L$ and $R$ such that no edge joins two vertices in the same set. In this paper, we denote a bipartite graph as $G=( L, R, E)$. Given a vertex $u\in G$, we use $N(u,G)$ to denote its neighbours in $G$ and $d_{max}$ to denote the maximum degree of $G$.

\noindent\textbf{Core number}~\cite{batagelj2003m}. The core number of a vertex $u$ in $G$, denoted by $core(u,G)$, is the largest possible integer $c$ such that there exists a subgraph $H\subseteq G$ containing $u$ and $min$ $\{|N(v,H)$ $||v\in H \}$ $\ge$ $c$.

\noindent\textbf{Degeneracy}. The maximum core number of $G$ is also called the degeneracy of $G$, denoted as $\delta(G)$.


\noindent\textbf{Biclique}. Given $G$, a pair of vertex sets $(A\subseteq L,B \subseteq R)$ is a biclique if  $\forall$$(u,v) \in A\times B $, $(u,v)$ $\in E$.

For instance, given the bipartite graph shown in Figure~\ref{fig:example}(b), $(\{1,2\},\{7\})$, $(\{3,4,5\},\{9,10\})$ induced subgraphs are bicliques.

\noindent\textbf{Balanced biclique}. A biclique $(A,B)$ is a balanced biclique, if $|A|=|B|$.

For instance, balanced bicliques in Figure~\ref{fig:example}(b) include $(\{1\},\{7\})$, $(\{2\},\{8\})$, $(\{3,4\},\{9,10\})$, etc.

\begin{rproblem}\textbf{Maximum balanced biclique problem}.\label{p:1} \textit{Given a bipartite graph $G=$($L,R,E$), find a balanced biclique $(A,B)$ such that there is no other balanced biclique $(A',B')$ with $|A'|+|B'|> |A|+|B|$.}
\end{rproblem}

For instance, $(\{1,2\},\{6,7\})$ is a maximum balanced biclique for the bipartite graph shown in Figure~\ref{fig:example}(a). Similarly, $(\{3,4\},\{9,10\})$ is a maximum balanced biclique for the bipartite graph shown in Figure~\ref{fig:example}(b).

\section{State of the Art}\label{sec:sota}

In this section, we revisit the state-of-the-art algorithm \cite{ZHOU2018834}, denoted by \textsf{ExtBBClq}, for solving the \textsf{MBB} problem exactly.

\textsf{ExtBBClq} is a branch and bound algorithm with upper bound based pruning.
The branch and bound part is inspired by the enumeration proposed in \cite{10.1007/978-3-319-07046-9_16}, which is an adaption from the maximal clique enumeration algorithm~\cite{10.1145/362342.362367}.
The algorithm starts the branch and bound procedure for enumerating all bicliques with vertices in non-increasing order according to their global degrees.
To efficiently compute an \textsf{MBB}, when branching at a vertex, an upper bound estimation is applied to prune non-promising branches.

The upper bounds used in~\cite{ZHOU2018834} are summarized below.
Given a vertex $v\in L$ , its upper bound is defined as the largest integer $i_{v}$ such that there are $i_{v}$ vertices in $L$ where each of the vertices has at least $i_{v}$ common neighbours with $v$.
The upper bound for a vertex in $R$ is defined similarly.
Therefore, given a vertex $u\in L\cup R$, its tight upper bound is defined as the largest integer $t_{u}$ such that there exists $t_{u}$ vertices in $N(u,G)$ with upper bound at least $t_{u}$.
The upper bound for every vertex is precomputed due to the high time complexity of computation.
When branching at $v$, if $2$ $\times$ $t_{v}$ is less than the maximum balanced biclique found so far, this branch is pruned.

In this paper, we use \textsf{ExtBBClq} as one of our baselines.
In fact, \textsf{ExtBBClq} essentially reduces the \textsf{MBB} problem to the maximal biclique enumeration (MBE) problem. We build several baselines using other state-of-the-art MBE algorithms with upper bound based prunings. Details are shown in the experimental studies.

\textsf{ExtBBClq} suffers from several shortcomings.
For dense bipartite graphs, the upper bound based pruning is less effective because every vertex looks promising according to their tight upper bounds. For instance, given the bipartite graph in Figure~\ref{fig:example}(a), every vertex has a looser upper bound of no less than $6$ whereas the size of an exact \textsf{MBB} is $4$.
For sparse bipartite graphs, the applied total search order has limited effectiveness for finding a large result at an early stage of the search, which limits the pruning effectiveness.
Besides, the total search order cannot tightly bound the search space, which results in high time complexity of \textsf{ExtBBClq}.

Bearing the above shortcomings in mind, we propose novel and efficient algorithms for dense and sparse bipartite graphs. 

\section{A Novel Algorithm for Dense Bipartite Graphs}\label{sec:dense}
Efficiently searching an \textsf{MBB} in dense bipartite graphs is very important.
There are two cases: 1) the input bipartite graph of an application is dense itself; 2) the original sparse bipartite graph of an application may be pruned and the remaining subgraphs become dense.
In both cases, a fast algorithm dedicated for dense bipartite graphs is the key for speeding up the search.

In this section, we propose a novel reduction, branch and bound algorithm, denoted by \textsf{denseMBB}, for those bipartite graphs that are sufficiently dense, where $|E|$ is $|L|\times |R| \times 80\%$ at least for a given bipartite graph $G=(L,R,E)$. As discussed, for real applications such as VLSI design, dealing with bipartite graphs with such high density is very common.

\noindent\textbf{Idea of our approach}.
We find that when a bipartite graph is sufficiently dense, the \textsf{MBB} problem can be solved in polynomial time.
As such, when branching at a vertex, we propose a branching strategy that aims to branch at a vertex which makes the remaining subgraphs denser and polynomially solvable as soon as possible.
Given the fact that the input graph is dense, the search can approach polynomially solvable subgraphs quickly using the above branching strategy, which substantially increases the performance.
Moreover, when an input graph is sufficiently dense, it is polynomial time solvable directly.

\subsection{Basic Enumerations}
In this section, we show the enumeration scheme that we use.
We explain it here since it is different from the existing works and it is fundamental for the correctness proof and the time complexity analysis of our advanced approach.

Algorithm~\ref{alg:basicbb} shows the enumeration scheme. It works on three pairs of sets, denoted as $(A,B)$, $(C_{A}, C_{B})$, and $(A^{*},B^{*})$.
$(A,B)$ is for storing the intermediate result of a balanced biclique.
$(C_{A},C_{B})$ contains candidate vertices to further expand $(A,B)$.
$(A^{*},B^{*})$ stores the \textsf{MBB} found so far.
Initially, the sets in $(A,B)$ and $(A^{*},B^{*})$ are empty while $C_{A}=L$, $C_{B}=R$ for $(C_{A}, C_{B})$.

\begin{algorithm}[t]
\tcc{bounding: simple pruning}
\textbf{if} it satisfies the bounding condition \textbf{then} \textbf{return}\; \label{basicBB:prune}

\tcc{checking maximality}
\If{$C_{A}$ == $\emptyset$}{
    \If{$min\{|A|,|B|\}$ $>$ $|A^{*}|$}{
        make (A,B) balance\;
       $(A^{*}, B^{*})$ $\leftarrow$ $(A, B)$\; \tcp{\scriptsize $(A^{*},B^{*})$ is a global variable}
    }
}
\tcc{branching}

Select a vertex from $u$ from $C_{A}$\;
\textsf{basicBB}$((B, A\cup \{u\}), (C_{B}\cap N(v,H),C_{A}\setminus\{u\}))$\label{basicbb:switch}\;
\textsf{basicBB}$((A,B), (C_{A}\setminus\{u\}, C_{B}))$\label{basicbb:switch2}\;

\caption{\textsf{basicBB($(A,B), (C_{A},C_{B})$)}}\label{alg:basicbb}
\end{algorithm}

Algorithm~\ref{alg:basicbb} finds an \textsf{MBB} via a search space that is a binary tree (lines \ref{basicbb:switch} and \ref{basicbb:switch2}). It is efficient for enumerating balanced bicliques because of the following reasons.
Firstly, it only considers vertices that can formulate bicliques with $(A,B)$, which is ensured by set operations in line~\ref{basicbb:switch}.
Secondly, the bicliques enumerated by Algorithm~\ref{alg:basicbb} are near balanced, i.e., the difference between $|A|$ and $|B|$ is no more than $1$. This is because the recursive calls switch the inputs, ensuring $A$ and $B$ are enlarged in turn. As such, Algorithm~\ref{alg:basicbb} avoids enumerating a large number of imbalanced bicliques while finding an \textsf{MBB}.

Algorithm~\ref{alg:basicbb} also applies intuitive prunings (line~\ref{basicBB:prune}). Given a recursion with $(A,B)$, $(L,R)$ and $(A^{*},B^{*})$, this recursion can be terminated if the following \textbf{bounding condition} is satisfied: $2\times min\{|A|+|L|, |B|+|R|\}$ $<$ $|A^{*}|+|B^{*}|$.
The correctness of the bounding condition is obvious. It indicates that the remaining search space cannot hold any balanced biclique with size greater than $(A^{*},B^{*})$.

Due to the simplicity of Algorithm~\ref{alg:basicbb}, we omit the correctness proof. The time complexity of Algorithm~\ref{alg:basicbb} is $\mathcal{O}^{*}(2^{n})$.

Next we will propose novel techniques that reduce $\mathcal{O}^{*}(2^{n})$ to $\mathcal{O}^{*}(1.3803^{n})$ and devise prunings that make $1.3803^{n}$ near constant for dense bipartite graphs.

\begin{figure}
\center
    \subfloat[Odd path]{\includegraphics[width=2.4cm]{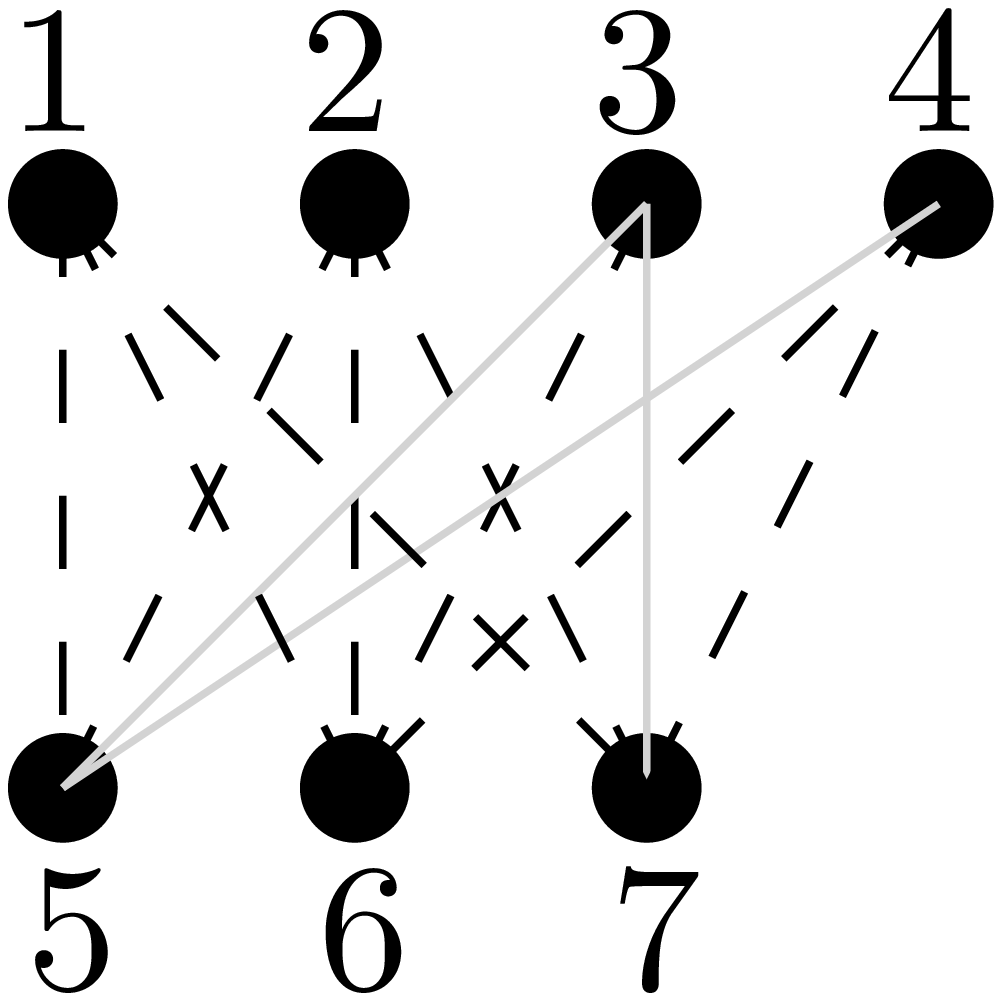}}
    \subfloat[Even path]{\includegraphics[width=2.3cm]{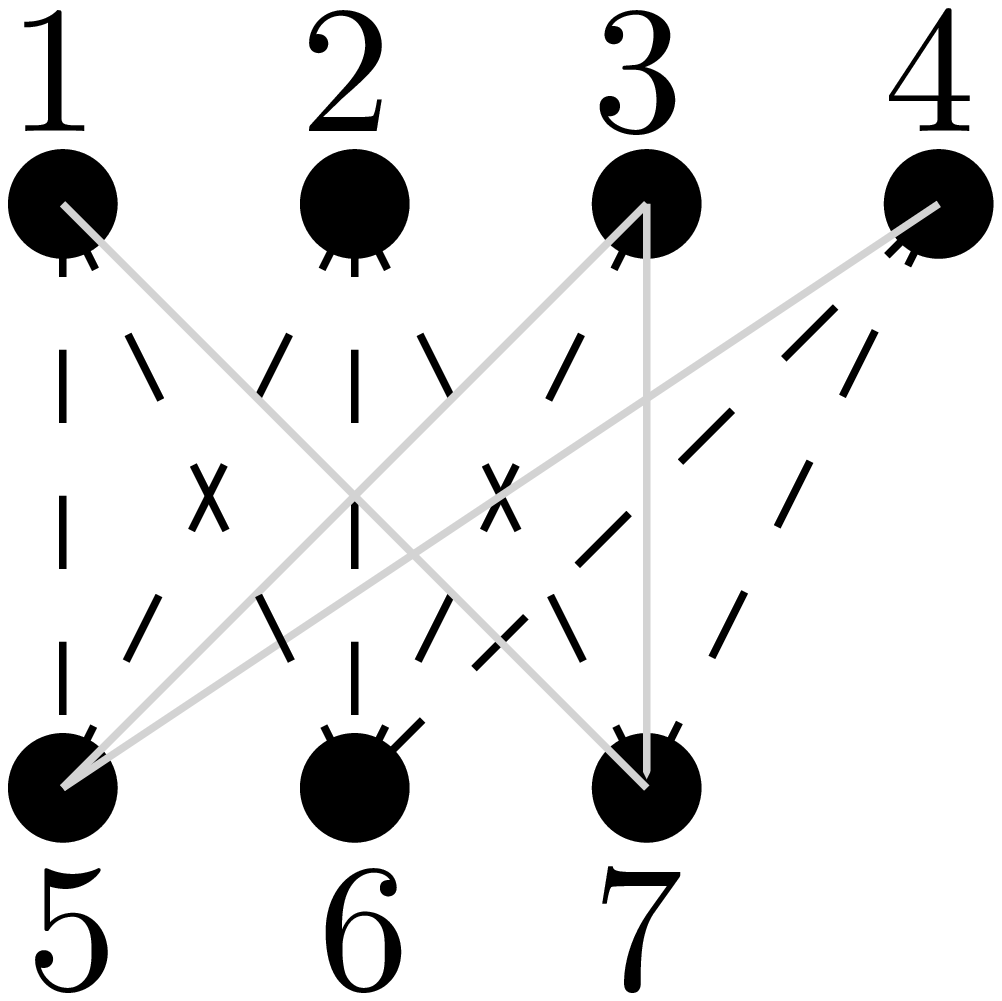}}
	\subfloat[Cycle]{\includegraphics[width=2.3cm]{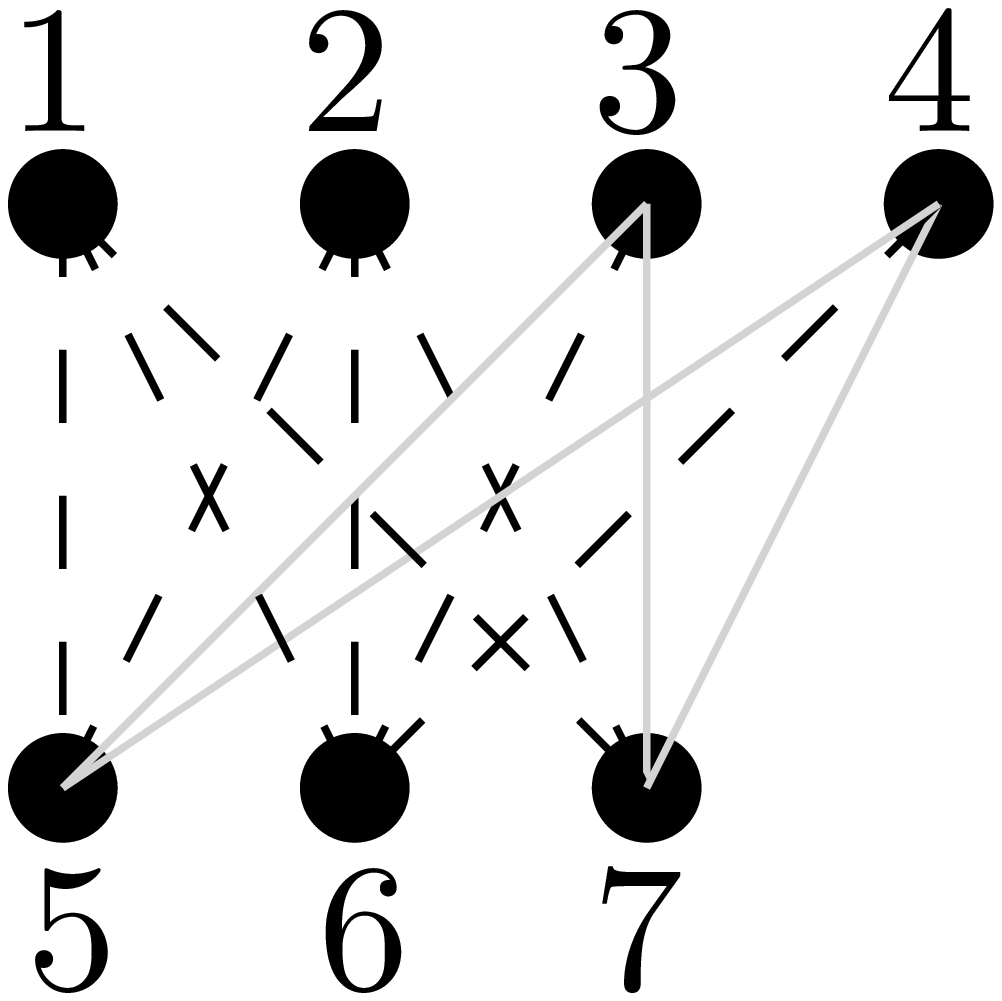}}
	\caption{Examples for polynomially solvable cases}\label{fig:pcases}
\vspace{-15pt}
\end{figure}

\subsection{Optimizations for Dense Bipartite Graphs}

In this section, we introduce the findings that help us improve the basic enumeration, leading to a novel reduction, branch and bound algorithm.

\noindent\textbf{Reduction}. We start the optimization with two simple but effective reduction rules which are applied for every recursion with $(A,B)$, $(C_{A},C_{B})$ and $(A^{*},B^{*})$ if possible.

\begin{lemma}\label{le:allc} \textbf{All connection reduction rule}.
Given a vertex $u\in C_{A}$ ($v\in C_{B}$), if $u$ ($v$) connects to every vertex in $C_{A}$ ($C_{B}$), move $u$ ($v$) from $C_{A}$ ($C_{B}$) to $A$ $(B)$.
\end{lemma}

\begin{lemma}\label{le:r1} \textbf{Low degree reduction rule}.
Given a vertex $u\in C_{A}$ ($v\in C_{B}$), if $deg(u,G(C_{A}\cup C_{B}))$ ($deg(v,G(C_{A}\cup C_{B}))$) is less than $|B^{*}|-|B|$ ($|A^{*}|-|A|$), remove $u$ ($v$) from $C_{A}$ ($C_{B}$).
\end{lemma}

The correctness of the above two reductions is clear. The reductions are applied until no vertices can be removed.

Next we show the techniques that lead to the algorithm with better complexity.
The intuition is that for any recursive call in Algorithm~\ref{alg:basicbb}, if we can guarantee that the two branches created would reduce the size of the candidate sets by at least $4$ and $1$ respectively, i.e., the worst branching factor~\cite{cormen2009introduction} is $(4, 1)$, the number of leaves of the recursion tree can be bounded by $\mathcal{O}(1.3803^{n})$. This can be achieved by: when a recursion will lead to worse branching factors (e.g. $(3,1)$), we do not continue the recursion but begin to solve the current sub-problem with a polynomial solution. 

\noindent\textbf{Polynomially solvable cases}.
We first introduce three definitions and then three important observations leading to polynomial time solvable cases.

\noindent\underline{\textit{Size constraint $(a,b)$ biclique problem}}. The size constraint $(a,b)$ biclique problem is defined as: given a bipartite graph $G=$($L,R,E$), and a pair of integers $(a,b)$, determine if there is a biclique $(A,B)$ in $G$ such that $|A|\ge a$ and $|B|\ge b$. We abbreviate size constraint $(a,b)$ biclique problem as $(a,b)$ biclique problem.

\noindent\underline{\textit{Maximal instances of $(a,b)$ biclique problem}}. We name an instance of $(a,b)$ biclique problem as maximal $(a,b)$ biclique problem for a bipartite graph if there exists no size $(a',b')$ biclique in the bipartite graph such that $a'$,$b'$ satisfy one of the conditions: 1) $a'= a$ and $b'> b$, 2) $a'> a$ and $b'= b$, 3) $a'> a$ and $b'> b$.

\noindent\underline{\textit{Bipartite complementary graph}}. Given a bipartite graph $G=(L,R,E)$, its bipartite complementary graph is defined as $\overline{G}=(L,R,\overline{E})$, where $\overline{E}$ is $L\times R\setminus E$.

\begin{obs}\label{ob:p1}
    Given a bipartite graph $H$, if $\forall$$u$ $\in$ $L(H)$, $deg(u,H)$ $\ge$ $|R(H)|-2$ and $\forall$$v\in R(H)$ $deg(v,H)\ge |L(H)|-2$, then the non-trivial parts of the bipartite complementary graph of $H$ would be a combination of even length paths, odd length paths and cycles.
\end{obs}

\begin{obs}\label{ob:p2}
    Given a bipartite graph $P$  that is  an even length path, an odd length path or a cycle with length of $p$, the maximal instances of $(a,b)$ bicliques in the bipartite complementary graph of $P$ are determined, shown below:
    \begin{enumerate}

      \item odd length path ($|L(P)|=|R(P)|$): $(0,\frac{p+1}{2})$, $(1,\frac{p+1}{2}-1)$, $\ldots$, $(\frac{p+1}{2}-1,1)$, $(\frac{p+1}{2},0)$.
      \item even length path($|L(P)|\ne |R(P)|$): $(0,\frac{p}{2}+1)$, $(1,\frac{p}{2}+1-1)$, $\ldots,$ $(\frac{p}{2}+1-2,2)$, $(\frac{p}{2}+1-1,0)$ if $|L(P)|< |R(P)|$ and $(0,\frac{p}{2}+1-1)$, $(2,\frac{p}{2}+1-2)$, $\ldots$, $(\frac{p}{2}+1-1,1)$, and $(\frac{p}{2}+1,0)$ if $|L(P)|> |R(P)|$.

      \item cycle: $(0,\frac{p}{2})$, $(\frac{p}{2},0)$ and $(2,\frac{p}{2}-1)$, $(3, \frac{p}{2}-2)$, $\ldots$, $(\frac{p}{2}-1,2)$ for $p> 4$ and $(0,\frac{p}{2})$, $(\frac{p}{2},0)$ for $p=4$.
    \end{enumerate}
\end{obs}


\begin{obs}\label{ob:p3}
     Given a bipartite graph $P$  that is  an even length path, an odd length path or a circle with length of $p$, any instance of $(a,b)$ bicliques in the bipartite complementary graph of $P$ is polynomially solvable.
\end{obs}

\noindent\textbf{Example}. In Figure~\ref{fig:pcases}, we show some examples for the above observations. Edges with dashed lines are the real edges in the bipartite graphs whereas lines with grey colour denotes the edges that are in their bipartite complimentary graphs.

\textit{Odd path}. For Figure~\ref{fig:pcases}(a), its complimentary bipartite graph contains an odd path with length of $3$ (grey lines), all possible maximal biclique instances for the vertices in this path induced subgraph of Figure~\ref{fig:pcases}(a) are $(0,2)$, $(1,1)$, and $(2,0)$ bicliques.

\textit{Even path}. For Figure~\ref{fig:pcases}(b), its complimentary bipartite graph forms an even path with length of $4$ and $|\{1,3,4\}|>|\{5,7\}|$, all possible maximal biclique instances are $(0,2)$, $(1,1)$, and $(2,0)$ bicliques for the vertices of the path induced subgraph of Figure~\ref{fig:pcases}(b).

\textit{Cycle}. For Figure~\ref{fig:pcases}(c), its complimentary bipartite graph is a cycle, all possible maximal biclique instances are $(0,2)$, $(2,0)$ for the vertices in this cycle induced subgraph of Figure~\ref{fig:pcases}(c).

We are ready to give the lemma below.

\begin{lemma}\label{le:pcs}
Given a subgraph $H$, if $\forall$$u$ $\in$ $L(H)$, $deg(u,H)$ $\ge$ $|R(H)|-2$ and $\forall$$v\in R(H)$ $deg(v,H)\ge |L(H)|-2$, the \textsf{MBB} problem can be solved in polynomial time.
\end{lemma}

We use Algorithm~\ref{alg:dyMBB} that runs in polynomial time to show the correctness of Lemma~\ref{le:pcs}.

The intuitions of Algorithm~\ref{alg:dyMBB} can be summarized below. Given $H$ satisfying the conditions in Lemma~\ref{le:pcs}, the non-trivial part of $\overline{H}$ shall consist of a combination of odd paths, even paths, and cycles (Observation~\ref{ob:p2}).
Therefore, all possible maximal instances of $(a,b)$ bicliques of $H$ can be built by checking the combinations of the trivial part of $\overline{H}$ and different maximal instances of $(a,b)$ bicliques that exist in the bipartite complementary graphs of the odd paths, even paths, and cycles.
After knowing all possible maximal instances of $(a,b)$ bicliques of $H$, the size of the \textsf{MBB} can be easily derived and an \textsf{MBB} can be found easily.
We shall show that all possible maximal instances of $(a,b)$ bicliques that $H$ contains can be built efficiently using the dynamic programming technique.

To embed the algorithm with the basic enumeration and reduction rules, Algorithm~\ref{alg:dyMBB} works on a partial result $(A,B)$ and a candidate set pair $(C_{A}, C_{B})$ when $(C_{A}, C_{B})$ induced subgraph satisfies the conditions stated in Lemma~\ref{le:pcs}, and thus deemed as polynomially solvable.
Algorithm~\ref{alg:dyMBB} first initializes a table that contains all the possible instances of bicliques that may be an \textsf{MBB} (lines \ref{dyMBB:si} to \ref{dyMBB:ei}), where value $1$ indicates that initially the maximal instance exists.
Algorithm~\ref{alg:dyMBB} builds actual \textit{maximal} biclique instances by combining the current known biclique and a new biclique implied by a path/cycle subgraph via lines \ref{dyMBB:ls} to \ref{dyMBB:le}.
The values of corresponding cells denote how many components that are absorbed to build the current biclique, each of which comes from a path or cycle in the complementary graph.
Using these values significantly reduces the number of cells in the table to be evaluated in the next loop.
After the loop, all possible \textit{maximal} instances of $(a,b)$ biclique that can be derived from $(A,B)$ and $(C_{A}, C_{B})$ are marked as none-zero values.
The largest instance of \textsf{MBB} can be easily derived via line \ref{dyMBB:mi}.
Then a new \textsf{MBB} would be computed if it is greater than the best \textsf{MBB} found so far (lines \ref{dyMBB:ms} and \ref{dyMBB:me}).

\begin{algorithm}[t]
\tcc{initializing a table}
 create $t[|A|+|C_{A}|][ |B|+|C_{B}|]$\;\label{dyMBB:si}
 create a hash table $h(\cdot)$\;

$t[|A|][|B|]$ $\leftarrow$ 1, create a hash table $h(\cdot,\cdot)$ $h(|A|,|B|)$ $\leftarrow$ $\{\{(|A|,|B|),0)\}\}$,

\label{dyMBB:ei}
$\mathcal{P}$ $\leftarrow$ compute odd paths, even paths, and cycles shown in Observation~\ref{ob:p1}, where the first element of the set is indexed as $1$;

\tcc{deriving the sizes of maximal balanced bicliques}
\For{$p$ $=$ $1$ to $|\mathcal{P}|$}{\label{dyMBB:ls}
    \ForEach{t[i][j]==p}{
        \ForEach{maximal (a,b) biclique instance of $\mathcal{P}[p]$}{
            \If{$t[i+a][j+b]\ne p+1$}{
                $t[i+a][j+b]=p+1$\;
                 $h(i+a,j+b)$ $\leftarrow$ $h(i,j)$ $\cup$ $ \{(a,b),p+1\}$\; \label{dyMBB:le}
            }

        }
    }
}

\tcc{finding the largest size of balanced biclique}

$(i,j)\leftarrow$ $\argmax_{(i,j)}\{min\{i,j\}|c[i][j]\ne 0 \}$ \; \label{dyMBB:mi}

\If{$min\{i,j\} >|A^{*}|$}{\label{dyMBB:ms}
    Access how $(i,j)$ is built via $h(i,j)$\;
    $(A^{*},B^{*})$ $\leftarrow$ compute (i,j) biclique and make it balance \;
    \Return $(A^{*},B^{*})$\;\label{dyMBB:me}
}

\Return $(\emptyset,\emptyset)$\;

\caption{\textsf{dynamicMBB}($(A,B), (C_{A},C_{B})$)} \label{alg:dyMBB}
\end{algorithm}

Algorithm \ref{alg:dyMBB} correctly finds an \textsf{MBB} based on the discussion above.
Algorithm \ref{alg:dyMBB} runs in $\mathcal{O}(n^{2})$ clearly since in the worst case the loop (lines \ref{dyMBB:ls} to \ref{dyMBB:le}) accesses all the cells of $t$. In fact, it is much faster. Due to limited space, obvious prunings applied in Algorithm~\ref{alg:dyMBB} are not shown.


\noindent\textbf{Branching techniques}. According to \textsc{Lemma}~\ref{le:pcs}, branching at a vertex leading to polynomially solvable cases would lead to fast search.
Therefore, a simple branching strategy is always to branch at a vertex that misses greater than 2 neighbours.

\noindent\textbf{Discussion}. It is natural to ask whether we can use the well-studied missing neighbour reduction techniques \cite{10.1145/3292500.3330986}, typically for finding a maximum clique in dense general graphs, for speeding up \textsf{MBB} search for dense bipartite graphs. Unfortunately we could not transplant the techniques.
For the clique problem, it has linear size self-reducible property, i.e., if there is a size $k$-clique in $G$, then there must be a size $(k-1)$-clique for any $k\ge 2$ and the total number of self-reducible problems is up to $k$.
However, for the biclique problem, if there is an $(a,b)$ biclique in $G$, there would be $\mathcal{O}(a\cdot b)$ instances of $(a', b')$ biclique problems that can lead to a size $(a,b)$ biclique. As such, missing neighbour reduction techniques may not be able to simplify \textsf{MBB} search.

\begin{algorithm}[t]
\textbf{if} it satisfies the bounding condition \textbf{then} \textbf{return}\;
\tcc{reduction, Lemmas~\ref{le:allc} and \ref{le:r1}}
$((A,B), (C_{A},C_{B}))$ $\leftarrow$ reduce $(C_{A},C_{B})$ induced subgraph \;
\textbf{if} it satisfies the bounding condition \textbf{then} \textbf{return}\;
\tcc{solving special cases, Lemma~\ref{le:pcs}}
\If{$(C_{A},C_{B})$ induced subgraph is polynomially solvable}
{
   $(A,B)$ $\leftarrow$ \textsf{dynamicMBB}($(A,B),(C_{A},C_{B})$)\;
    \If{$|A|\ne \emptyset$}{

        $(A^{*}, B^{*})$ $\leftarrow$ $(A, B)$\; \tcp{\scriptsize $(A^{*},B^{*})$ is a global variable}
    }
    \Return \;
}

\tcc{branching according to Lemma~\ref{le:pcs}}

Select a vertex  $u$ from $C_{A}\cup C_{B}$ missing at least 3 neighbours in the $C_{A}\cup C_{B}$ induced subgraph \;
\eIf{$u\in C_{A}$}{
    \label{rbb:switch1}  \textsf{denseMBB}$((B, A\cup \{u\}), (C_{B}\cap N(u,H),C_{A}\setminus\{u\}))$\;
    \textsf{denseMBB}$((A,B), (C_{A}\setminus\{u\}, C_{B}))$\;
}{
    \label{rbb:switch2}  \textsf{denseMBB}$((B\cup \{u\}, A), (C_{B}\setminus\{u\},C_{A}\cap N(u,H)))$\;
    \textsf{denseMBB}$((A,B), (C_{A}, C_{B}\setminus\{u\}))$\;
}

\caption{\textsf{denseMBB($(A,B), (C_{A},C_{B})$)}}\label{alg:rbb}
\end{algorithm}

\subsection{The Algorithm}

Now we are ready to present the complete reduction, branch and bound algorithm.

\noindent\textbf{The algorithm}. The major steps are shown in Algorithm~\ref{alg:rbb}. Algorithm~\ref{alg:rbb} incorporates all the discussed theoretical findings to speed up \textsf{MBB} search: line $9$ is for optimizing branching, line $2$ is for reducing the subgraph as much as possible, lines $4$ to $8$ are for processing polynomially solvable cases whenever possible.

\noindent\textbf{Correctness}. Algorithms \ref{alg:rbb} correctly finds an \textsf{MBB}. In Algorithm \ref{alg:rbb}, the applied reductions prune fruitless vertices and once polynomially solvable cases are reached Algorithm \ref{alg:dyMBB} correctly solves them. 

\noindent\textbf{Time complexity}. The time complexity of Algorithm \ref{alg:rbb} is $\mathcal{O}^{*}(1.3803^{n})$. All polynomially solvable cases and the proposed branching strategy ensure that the worst branching factor is $(4,1)$, since $u$ is able to invalidate at least $3$ non-neighbors plus the removal of $u$ from the candidate set, resulting a reduction of at least $4$ vertices from the candidate sets. Therefore, the total number of recursion is bounded by $\mathcal{O}(1.3803^{n})$. For individual recursions, the time complexities are dominated by polynomially solvable cases, i.e., Algorithm \ref{alg:dyMBB}.
Therefore, the time complexity of Algorithm \ref{alg:rbb} is $\mathcal{O}(n^{2}\cdot 1.3803^{n})$, i.e., $\mathcal{O}^{*}(1.3803^{n})$.

We would like to highlight that for dense graph with the number of $|E|$ at the scale of $|L|\times |R| \times 80\%$ at least, Algorithm \ref{alg:rbb} most likely runs in $\mathcal{O}(n^{2})$ since it converges to polynomially solvable cases with near constant numbers of recursions. For example, the bipartite graph shown in Figure~\ref{fig:example}(a) can be solved in polynomial time directly since the bipartite graph directly satisfies the conditions in Lemma~\ref{le:pcs}.

\section{A Novel Algorithm for Large Sparse Bipartite Graphs}\label{sec:sparse}

Solving the \textsf{MBB} problem for large sparse bipartite graphs is important for applications such as biological data analysis. Existing \textsf{MBB} algorithms reduce the \textsf{MBB} problem to the maximal biclique enumeration problem with various of prunings.
As such, the time complexities of the existing algorithms cannot be better than that of the maximal biclique enumeration problem.
As far as we know, the state-of-the-art maximal biclique enumeration algorithm has the time complexity of $\mathcal{O}^{*}(n^{d_{max}})$, where $d_{max}$ is the maximum degree.
Given the fact that $d_{max}$ can be as high as $n$, applying our proposed Algorithm~\ref{alg:rbb} directly on large sparse bipartite graphs would lead to a better algorithm from theoretical perspective.
In fact, we can do much better.

In this section, we propose a novel \textsf{MBB} algorithm for large sparse bipartite graphs.

\begin{algorithm}[t]
\tcc{step 1: Heuristically find a large size balanced biclique and reduce $G$ as much as possible}

    $(A^{*},B^{*}), G' \leftarrow$ \textsf{hMBB}$(G)$\;

\tcc{step 2: Prepare locally dense subgraphs and refine the found large balanced biclique}

    $(A^{*},B^{*}), \mathcal{H}$ $\leftarrow$ \textsf{bridgeMBB}$((A^{*},B^{*}),G')$\;

\tcc{step 3: verify and derive MBB}

    $(A^{*},B^{*})$ $\leftarrow$ \textsf{verifyMBB}$((A^{*},B^{*}),\mathcal{H})$ \;
\textbf{return} $(A^{*},B^{*})$ \;
\caption{Search framework \textsf{hbvMBB}$(G)$}\label{alg:gMBB}
\end{algorithm}

\subsection{Overview of the Algorithm}
Intuitions of our algorithm can be summarized below.

Firstly, we separate heuristics from the exhaustive search, which brings advantages.
We can apply advanced heuristics that have a higher chance to find a global \textsf{MBB} before exhaustive search.
This increases the effectiveness of the upper bound based pruning phenomenally.

Secondly, to efficiently perform the exhaustive search, we propose a novel technique that transforms the bipartite graph that cannot be pruned into small but dense subgraphs. A tighter upper bound for each vertex can be derived within their local dense subgraphs, which further reduces the search space.
Impressively, experimental results demonstrate that for lots of real sparse datasets ($12$ out of $30$), an \textsf{MBB} can be derived without exhaustive search using the above techniques.

Last but not the least, we apply the proposed Algorithm~\ref{alg:rbb} to small but dense subgraphs that cannot be pruned. Due to the fact that remaining subgraphs exhibit high density and small size, this step takes near polynomial time in practice.

To effectively apply the aforementioned ideas, we propose a search framework consisting of three major steps, shown in Algorithm~\ref{alg:gMBB}.
The first step is for finding a large-size result heuristically and reducing the graph as much as possible.
The second step is for generating locally dense subgraphs. This step would further refine the maximality of the found \textsf{MBB} and prune the bipartite graph if possible.
The third step is for verifying the maximality of the found results.


Next, we expand each step in great detail.






\subsection{Heuristic and Reduction}\label{sec:handr}
In this section, we propose a fast heuristic \textsf{MBB} search algorithm, denoted by \textsf{hMBB} for effectively pruning sparse bipartite graphs.
Different from existing heuristic \textsf{MBB} search algorithms aiming for discovering a large \textsf{MBB} within reasonable time, as a subroutine of exact \textsf{MBB} search, we have the expectations for \textsf{hMBB} below.
Firstly, \textsf{hMBB} should be extremely fast, i.e., near linear time w.r.t. the size of a bipartite graph.
Secondly, \textsf{hMBB} should reduce the graph size as much as possible.

Now, we introduce the \textsf{hMBB} algorithm.


\noindent\textbf{Heuristics and reduction based approach}. \textsf{hMBB} follows heuristics and reduction pattern.
Let $(A^{*},B^{*})$ denote the maximum balanced biclique found so far, e.g., by a greedy algorithm, we apply reductions used in~\cite{ZHOU201986,wang2018new} below.

\begin{lemma}\label{le:corep}
    Given $(A^{*},B^{*})$, $\forall$$v$ not in $|A^{*}|+1$ core subgraph, $v$ cannot be a part of balanced biclique having size greater than $(A^{*},B^{*})$.
\end{lemma}


\noindent\textbf{The \textsf{hMBB} algorithm}.
\textsf{hMBB} is shown in Algorithm~\ref{alg:hMBB}. In the first step, Algorithm~\ref{alg:hMBB} first endeavours to find a large-size balanced biclique using the maximum degree based greedy rule and then applies a reduction based on Lemma~\ref{le:corep}. Due to the simplicity of the greedy algorithm, we omit its details.
After that, maximum core number based greedy rule is used to find a large-size balanced biclique and \textsc{Lemma}~\ref{le:corep} based reduction is applied again if a larger balanced biclique is found.

\begin{algorithm}[t]
$(A^{*},B^{*})$ $\leftarrow$ $(\emptyset,\emptyset)$\;
\tcc{degree based heuristic}
$(A,B)$ $\leftarrow$ compute a balanced biclique using maximum degree based heuristic\;
$(A^{*},B^{*})$ $\leftarrow$ $(A,B)$\;
\tcc{reduction}
reduce the graph to $G'$ using \textsc{Lemma}~\ref{le:corep} and then compute the degeneracy, core number for $G'$\;
\tcc{degeneracy based heuristic}

\textbf{if}$(2\delta(G'))==(|A^{*}|+|B^{*}|)$ \textbf{then} \textbf{return} $\{(A^{*},B^{*}),\emptyset\}$ \;
$(A, B)$ $\leftarrow$ compute a balanced biclique using maximum core number based heuristic \;

\tcc{reduction}
\If{$|A|$ $>$  $|A^{*}|$}{
    $(A^{*},B^{*})$ $\leftarrow$ $(A, B)$\;
    $G''$ $\leftarrow$ reduce the graph using \textsc{Lemma}~\ref{le:corep}\;
	compute the degeneracy of $G''$\;
    \textbf{if}$(2\delta(G''))==(|A^{*}|+|B^{*}|)$ \textbf{then} \textbf{return} $\{(A^{*},B^{*}),\emptyset\}$ \;
}

\textbf{return} $\{(A^{*},B^{*}), G'' \}$ \;

\caption{\textsf{hMBB$(G)$}}\label{alg:hMBB}
\end{algorithm}

\noindent\textbf{Early termination}.
We propose an early termination condition to avoid unnecessary further vertex deletions once an \textsf{MBB} is found. 

\begin{lemma}\label{le:ubp}
Let $(A^{*},B^{*})$ denote the maximum balanced biclique found up to the time in $G$, if $|A^{*}|+|B^{*}|$ equals to twice of the core number of $G$, then we can terminate the algorithm.
\end{lemma}


\noindent\textbf{Time complexity}. The time complexity of \textsf{hMBB} is $\mathcal{O}(|E|)$, which is dominated by the computation of core decomposition.
We may apply \textsf{hMBB} for top-$r$ maximum degree (core number) vertices and see if we can get a larger-size balanced biclique and further reduce the bipartite graph accordingly.

\noindent\textbf{Example}. Assume the graph shown in Figure~\ref{fig:example}(b) is the input for Algorithm~\ref{alg:hMBB}. Using degree based heuristic, it will find a size-$2$ balanced biclique. Using the core-based heuristic, where the core for each vertex is shown in Table~\ref{tab:core}, it will find a size-$4$ balanced biclique $(\{3,4\},\{9,10\})$. Using this result, Algorithm~\ref{alg:hMBB} detects that the condition in Lemma~\ref{le:ubp} is satisfied. Therefore, $(\{3,4\},\{9,10\})$ is the optimum result.

\subsection{Bridging to Maximality}
In this section, we propose techniques for preparing maximality verification.
We propose an approach that effectively transforms the residual subgraphs output by step $1$ into small but dense subgraphs without loss of global optimum results.
After the transformation, tighter upper bound for each vertex could derived, which would further prune subgraphs that are fruitless.

We first show theoretical findings that help us explain the above techniques.

\subsubsection{Measuring Bisparsity}
\begin{figure}
\center
	\subfloat[vertex 2]{\includegraphics[height=2cm]{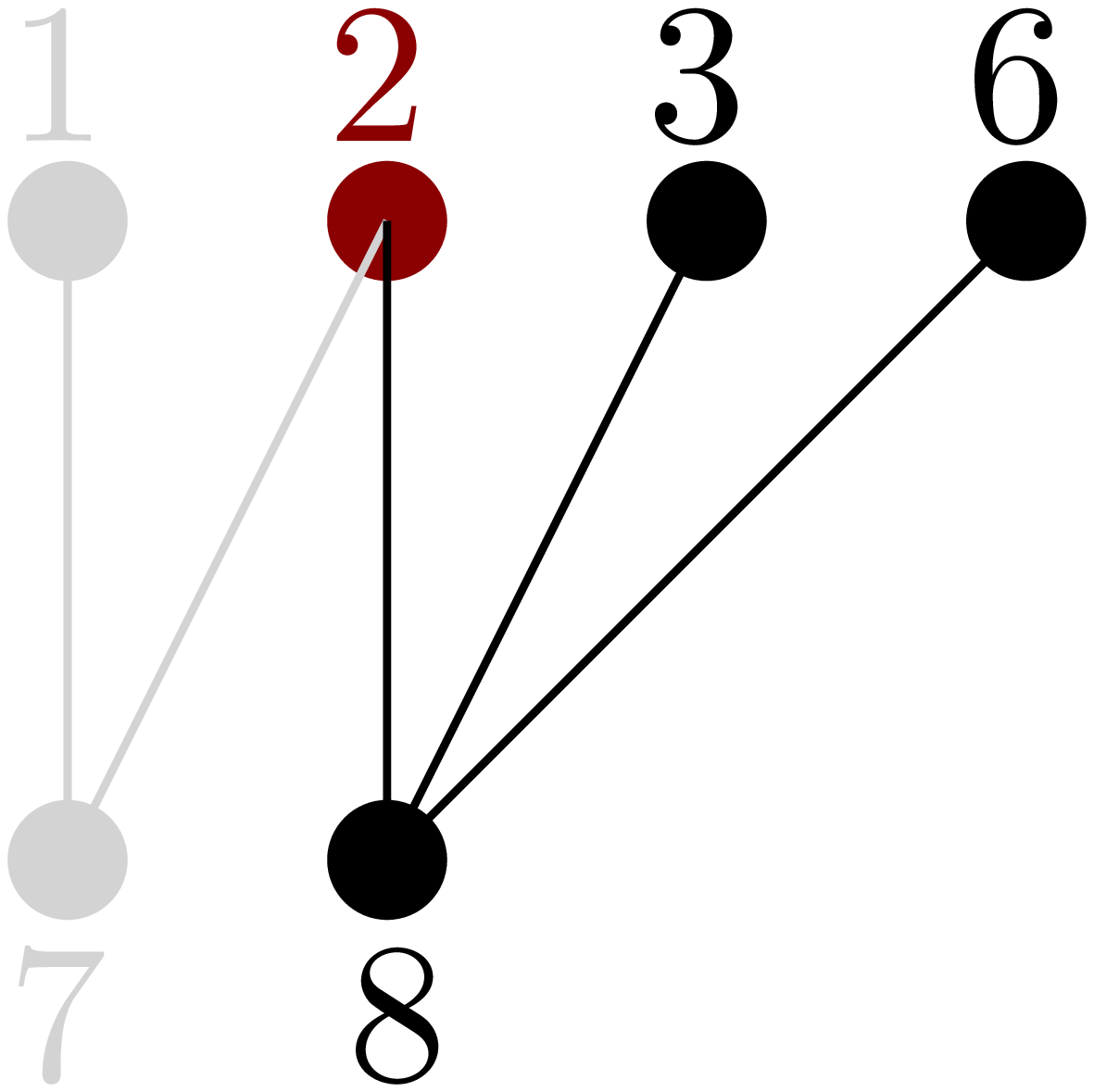}}
	\hspace{10pt}
	\subfloat[vertex 3]{\includegraphics[height=2cm]{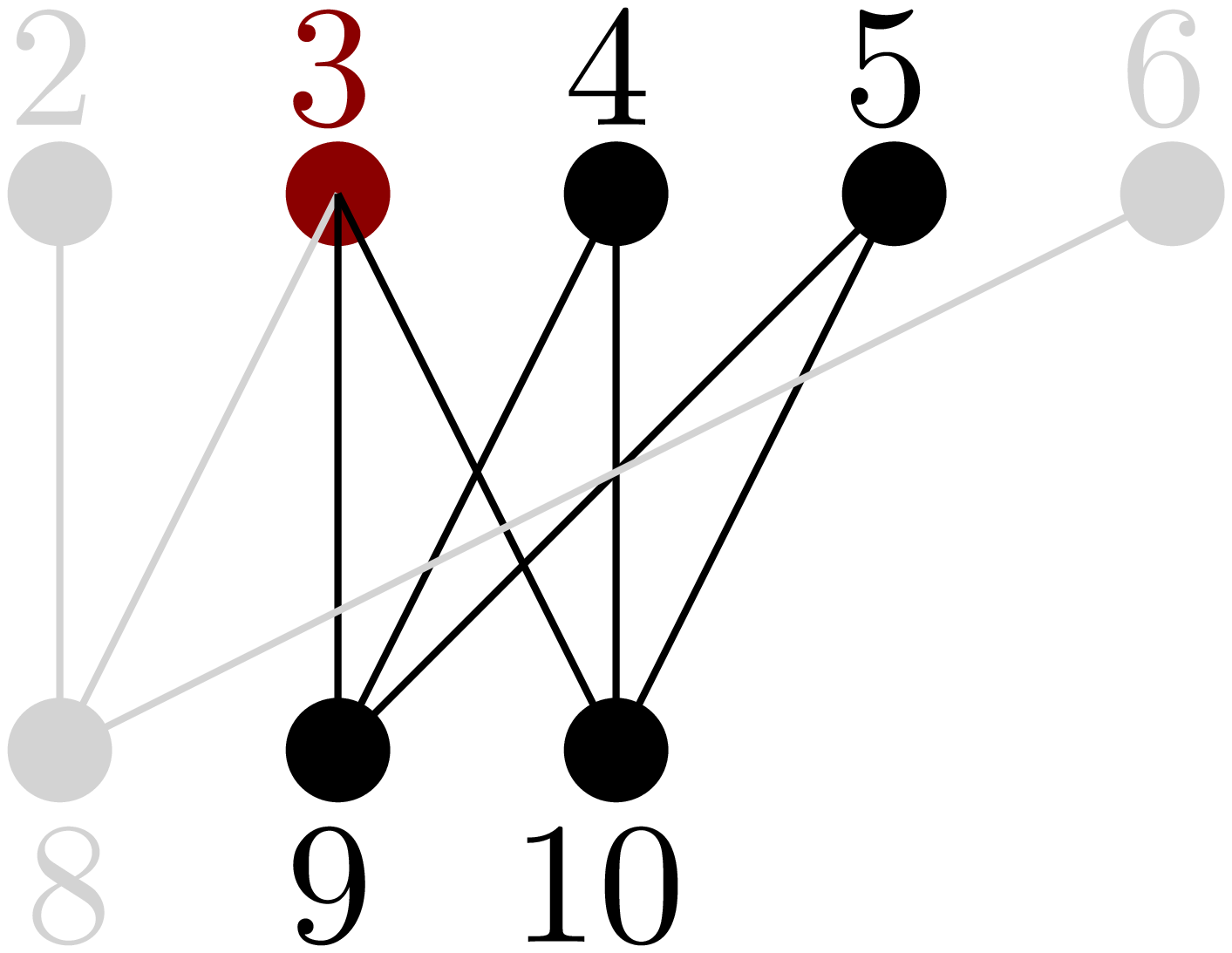}}
	\caption{$N_{\le 2}(\cdot,\cdot)$ and vertex centered subgraphs }\label{fig:orderGraph}
\vspace{-10pt}
\end{figure}

For general graphs, the sparsity measurement, known as degeneracy, is derived based on $1$-hop neighbours of every vertex.
Degeneracy lays the foundation for designing efficient algorithms for the maximum clique problem for general graphs~\cite{10.1145/3292500.3330986}.
A general graph with degeneracy of $\delta$ would bound the enumeration depth to $\delta$ for the clique problem, which is significantly less than $d_{max}$ when the graph is sparse.

Different from general graphs, biclique problems for a bipartite graph have to consider both $L$ and $R$ sides.
To effectively capture the bipartite sparsity of a bipartite graph, we have to consider both sides. Therefore, we propose bisparity, which is measured by  bidegeneracy (bipartite degeneracy) derived from both $1$-hop neighbours and $2$-hop neighbours of every vertex in a bipartite graph.
Using bidegeneracy, we would show its effectiveness for reducing the search space for solving the \textsf{MBB} problem in large sparse bipartite graphs later.

We first formally define 2-hop neighbours.

\begin{definition} \textbf{2-hop neighbours}.
Given a vertex $u$ in $G$, we use $N_{2}(u,G)$ to denote the set of vertices who have the lengths of shortest path to $u$ at $2$ exactly.

\end{definition}
For instance, for vertex $2$ in Figure~\ref{fig:example}(b), the $2$-hop neighbours of $2$ are $\{1,3,6\}$.

\begin{definition} \textbf{$N_{\le 2}(u,G)$}.
We define $N_{\le 2}(u,G)$ to be the neighbours and $2$-hop neighbours of $u$ in $G$, i.e., $N_{\le 2}(u,G)$ $=$ $N(u, G)$$\cup$$N_{2}(u,G)$.
\end{definition}

For instance, for vertex $2$ in Figure~\ref{fig:example}(b), $N_{\le 2}(2,G)$ includes $\{1,3,6,7,8\}$.

Below, we propose novel definitions, bicore and bidegeneracy, which are fundamental for devising our proposed algorithm for large sparse bipartite graphs.

\begin{definition}\label{def:bicore} \textbf{Bicore number}.
Given a bipartite graph $G$, the bicore number of a vertex $u$, denoted by \textsf{bc}$(u)$, is the largest possible integer $k$ such that there exists a subgraph $H$ containing $u$ whose $min$ $\{|N_{\le 2}(v,H)||v\in H\}$ is $k$.
\end{definition}

\begin{definition}\textbf{Bidegeneracy}.
The maximum bicore number of $G$, is defined as the bidegeneracy of $G$, denoted as $\ddot{\delta}(G)$.
\end{definition}

Using bidegeneracy, a bidegeneracy order can be defined accordingly.

\begin{definition}\textbf{Bidegeneracy order}.
A permutation of $L\cup R$, \textsf{BDorder} $=(v_{1},\ldots,v_{|L|+|R|} )$ is a bidegeneracy order if every vertex $v_{i}$ has the smallest $|N_{\le 2}(v_{i})|$ in the subgraph of $G$ induced by $\{v_{i},\ldots, v_{|L|+|R|}\}$.
\end{definition}

\begin{table}
	\caption{Core and bicore number for Figure 1(b)}\label{tab:core}
	\begin{tabular}{|c|c|c|c|c|c|c|c|c|c|c|c|c|}
		\hline
		vertex & 1	&2	&3	&4	&5	&6	&7	&8	&9	&10	&11	&12\\
		\hline
		core($\cdot$)	&1&1&2&2&2&1&1&1&2&2&1&1\\
		\hline
		bc($\cdot$)	&2&3&4&4&4&3&2&3&4&4&3&3\\
		\hline
	\end{tabular}
\end{table}

\noindent\textbf{Example}. The bicore number for each vertex in Figure~\ref{fig:example}(b) is demonstrated in Table~\ref{tab:core}.
The bidegeneracy of the graph shown in Figure~\ref{fig:example}(b) is 4.
One of the bidegeneracy order for the graph in Figure~\ref{fig:example}(b) is \textsf{BDorder}$=$ $(1,7,2$$,6,11,$ $12$$,8,3,4,5,9,10)$.

\subsubsection{From Sparse to Dense}

We are ready to show an effective method for transforming residual subgraphs, output by step 1, into small dense subgraphs.

We first introduce two observations below.

\begin{obs} \textbf{Biclique search scope for a vertex}. Given a vertex $u$ of a bipartite graph $G$, all the bicliques where $u$ is involved are restricted within $u$ and $N_{\le 2}(u, G)$ induced subgraphs.
\end{obs}

For instance, for vertices $2$ and $3$ in Figure~\ref{fig:example}(b), the $2$ and $N_{\le 2}(2, G)$ induced subgraph is shown in Figure~\ref{fig:orderGraph}(a) and the $3$ and $N_{\le 2}(3, G)$ induced subgraph is shown in  Figure~\ref{fig:orderGraph}(b). Please ignore the colour differences in this example. Clearly, all maximal bicliques involving $2$ and $3$ are contained in Figures~\ref{fig:orderGraph}(a) and (b) respectively.

\begin{obs} \textbf{Total search order}. In general, given a bipartite graph, an exhaustive search of visiting all maximal balanced bicliques would follow a certain total search order of vertices $o=(v_{1},\ldots, v_{|L|+|R|})$. Following the total search order, when processing $v_{i}$, the exhaustive search only considers bicliques that must contain $v_{i}$ in $\{v_{i+1},\ldots, v_{|L|+|R|}\}$ induced subgraphs of $G$ to avoid duplicate combinations.
\end{obs}

Based on the above two observations, we can transform a bipartite graph into at most $|L|+|R|$ number of subgraphs, where each subgraph is defined below.

\begin{definition} \textbf{Vertex centred subgraph}.
Given a total search order $o =(v_{1},\ldots, v_{|L|+|R|})$ for a graph $G$ and $v_{i}\in o$, $v_{i}$ centred subgraph is defined as $v_{i}$ and $N_{\le 2}(v_{i},G)$ $\cap$ $\{v_{i+1},\ldots, v_{|L|+|R|}\}$ induced subgraph.
\end{definition}

For instance, given the bidegeneracy order of the graph in Figure~\ref{fig:example}(a), \textsf{BDorder}$=$ $(1,7,2$$,6,11,$ $12$$,8,3,4,5,9,10)$, vertices 3 and 4 centred subgraphs are black and red parts of Figures~\ref{fig:orderGraph}(a) and (b) respectively.
\begin{algorithm}[t]

Compute the bidegeneracy of $G'$\;

$\mathcal{H}$ $\leftarrow$ compute vertex centred graphs based on the bidegeneracy order\;
\ForEach{$H$ $\in$ $\mathcal{H}$ }{
    \If{$min(|L(H)|,R(H))$ $<$ $|A^{*}|$}{
        remove $H$ from $\mathcal{H}$\;
        \textbf{continue}\;
    }
    Compute the degeneracy for $H$\;
    \If{$\delta(H)\le |A^{*}|$}{
        remove $H$ from $\mathcal{H}$\;
        \textbf{continue}\;
    }
    $(A, B)$ $\leftarrow$ compute an \textsf{MBB} using the local core number based heuristic\;
    \If{$|A^{*}|< |A|$}{
        $(A^{*},B^{*})$ $\leftarrow$ $(A,B)$\;
    }
}

\textbf{return} $\{(A^{*},B^{*}), \mathcal{H} \}$ \;
\caption{\textsf{bridgeMBB}$((A^{*},B^{*}), G')$}\label{alg:bridge}
\end{algorithm}

\noindent\textbf{A search order for tightening search space}. We want to find an order that can tightly bound the total size of the vertex centred subgraphs. We show our findings below.

\begin{lemma}\label{le:degO}
Using non-increasing degree order, the total size of vertex centred subgraphs for $G$ is $\mathcal{O}((|L|+|R|)d_{max}^{2})$, where $d_{max}$ is the maximum degree of $G$.
\end{lemma}

\begin{lemma}\label{le:degenO}
Using degeneracy order, the total size of vertex centred subgraphs for $G$ is $\mathcal{O}((|L|+|R|)\delta(G)d_{max})$.
\end{lemma}

\begin{lemma}\label{le:bdO}
Using bidegeneracy order, the total size of vertex centred subgraphs for $G$ is $\mathcal{O}((|L|+|R|)\ddot{\delta}(G))$.
\end{lemma}

In real graphs, $\ddot{\delta}(G)$ is significantly smaller than $d_{max}$. Therefore, using bidegeneracy order would have much tighter bound. We will show how dense of a vertex centred subgraph is in the experimental studies.



\subsubsection{The Algorithm}

Based on the discussed theoretical findings, we propose Algorithm~\ref{alg:bridge} for step $3$ in Algorithm~\ref{alg:gMBB}.
It first computes bidegeneracy for the pruned graph ($G'$).
Then it generates vertex centred subgraphs using bidegeneracy order.

To prune the vertex centred subgraphs as much as possible, for each subgraph $H$, Algorithm~\ref{alg:bridge} applies prunings according to the size of $H$, degeneracy of $H$, and the maximum balanced biclique found up to the time.
The local upper bound for each subgraph is significantly improved.
Therefore, the pruning effectiveness is phenomenal.
For the subgraph $H$ that cannot be pruned, Algorithm~\ref{alg:bridge} applies maximum degeneracy based greedy algorithm, which attempts to find a larger \textsf{MBB} for maximizing pruning effects.

Algorithm~\ref{alg:bridge} returns the maximum balanced biclique found so far and vertex centred subgraphs that cannot be pruned.

\noindent\textbf{Time complexity}. The time complexity of Algorithm~\ref{alg:bridge} is dominated by bicore decomposition.
We propose Lemma~\ref{le:tbdec} to show the time complexity of bicore decomposition.

\begin{lemma}\label{le:tbdec}
	Given a bipartite graph $G$, there exists an algorithm that can perform  bicore decomposition for $G$ with time complexity of $\mathcal{O}(\sum_{u\in G} |N_{\le 2}(v,G)|)$.
\end{lemma}

\begin{algorithm}[t]
ordered list  $o\leftarrow \emptyset$\;
compute $N_{\le 2}(u,G)$	 for every $u$ in $G$\;
order the vertices in $G$ in increasing order according to $|N_{\le 2}(u,G)|$ for every $u$ \;
order the vertices in $G$ where every $u$ has the same $|N_{\le 2}(u,G)|$ in increasing order according to their neighbours \;
\ForEach{$u$ $\in$ $G$ in the order}{
	$bc[u]$=$N_{\le 2}(u,G)$\;
	append $u$ at the end of $o$\;
	\ForEach{$v\in N_{\le 2}(u,G)$}{
		$N_{\le 2}(v,G)$ $\leftarrow$ $N_{\le 2}(v,G)\setminus \{u\}$\;
        adjust the position of $u$ according to the order of lines $3$ and $4$. 		
	}
}
\caption{Bicore decomposition}\label{alg:bicore}
\end{algorithm}

It is non-trivial to design an efficient peeling algorithm for bicore decomposition.
This is because when removing $u$, for every $v\in N_{\le2}(u,G)$, $|N(v,G)|$ may reduce more than $1$.
If the peeling order is not chosen carefully, we may need extra computation to check how many neighbours that each $v\in N_{\le2}(u,G)$ loses after removing $u$.
To avoid such pessimistic situation, we show our theoretical finding below.


\begin{lemma}\label{le:peeling}
When peeling, if every time the removed $u$ in $H$  satisfies two conditions: 1) $u$ has the minimum $|N_{\le 2}(u,H)|$ and 2) $u$ has the minimum $|N(u,H)|$ among all vertices satisfying condition 1), then for every vertex $v$ in $N_{\le 2}$ $(u,H)$, its $|N_{\le 2}(v,H)|$ shall reduce by no more than $1$.
\end{lemma}

The above lemma can be proved easily via contradiction.

Based on \textsc{Lemma}~\ref{le:peeling}, an $\mathcal{O}(\sum_{u\in G} |N_{\le 2}(v,G)|)$ bicore decomposition algorithm is shown in Algorithm~\ref{alg:bicore}.
To achieve $\mathcal{O}(\sum_{u\in G} |N_{\le 2}(v,G)|)$ time complexity, we adapt the bucket sort based approach~\cite{batagelj2003m}. We need a bucket for effectively maintaining an order according to the union of neighbours and 2-hop neighbours for speeding up the checking of condition 1) in \textsc{Lemma}~\ref{le:peeling}.
We also need a set of buckets for effectively maintaining a set of orders according to neighbours of vertices whose union of neighbours and $2$-hop neighbours have the same size for speeding up the checking of condition 2) in \textsc{Lemma}~\ref{le:peeling}.  As such, lines 3 and 4 can be done within $\mathcal{O}(|L|+|R|)$ and line 10 can be done in constant time.
The dominating parts are line 2 and lines 5 to 10 and each of them takes $\mathcal{O}(\sum_{u\in G} |N_{\le 2}(v,G)|)$.

\subsection{Maximality Verification}
In this section, we propose how to efficiently verify the maximality of the result found up to the time on the set of vertex centred subgraphs that cannot be pruned.

\noindent\textbf{Maximality verification algorithm}.
Algorithm~\ref{alg:mv} shows how to verify the maximality of $(A^{*}, B^{*})$.
It checks all vertex centred graphs that cannot be pruned by techniques that have been discussed.
For one vertex centred graph $H$, Algorithm~\ref{alg:mv} would first further reduce $H$ according to Lemma~\ref{le:ubp} if a larger \textsf{MBB} is found (line \ref{mv:p}).
After that, Algorithm~\ref{alg:mv} calls Algorithm~\ref{alg:rbb} to check whether the remaining $H$ contains a balanced biclique larger than $(A^{*}, B^{*})$ (lines \ref{mv:r1} to \ref{mv:r2}). 
If there is a larger one, $(A^{*}, B^{*})$ will be updated.
After checking all vertex centred graphs, Algorithm~\ref{alg:mv} returns the optimum result.







\subsection{Analysis}
In this section, we show the correctness of Algorithm~\ref{alg:gMBB} embedding all the proposed techniques and analyze its time complexity.

\noindent\textbf{Correctness}. The correctness of Algorithm~\ref{alg:gMBB} can be derived as follows.
Firstly, all the prunings are correct.
Secondly, the vertex centred subgraphs are generated based on a total search order, therefore, the subgraphs cover all possible bicliques that are promising.
Thirdly, all the promising vertex centred subgraphs are applied with exhaustive search (Algorithm~\ref{alg:rbb}) that is proven to be correct.
Therefore, Algorithm~\ref{alg:gMBB} is correct.


\noindent\textbf{Time complexity}. Algorithm~\ref{alg:gMBB} embedding all the proposed techniques finds an \textsf{MBB} in $\mathcal{O}^{*}($ $1.3803^{\ddot{\delta}})$.
The breaking down analysis is given below.
For step $1$ (Algorithm \ref{alg:hMBB}), the dominating computation is core decomposition that has time complexity of $\mathcal{O}(|E(G)|)$.
For step $2$ (Algorithms \ref{alg:bridge} and  \ref{alg:bicore}), the dominating computation is bicore decomposition that can be bounded by $\mathcal{O}$ $(\sum_{v\in G}$ $|N_{\le 2}$ $(v,G)|)$.
For step $3$ (Algorithm~\ref{alg:mv}), the time complexity is $\mathcal{O}(|V(G)|$ $\cdot$ $\ddot{\delta}^{2}$ $\cdot$ $1.3803^{\ddot{\delta}})$, since there are up to $|V(G)|$ number of vertex centred subgraphs that are evaluated by Algorithm~\ref{alg:mv}.
As such Algorithm~\ref{alg:gMBB} is dominated by step $3$, which is $\mathcal{O}^{*}(1.3803^{\ddot{\delta}})$.
In fact, Algorithm~\ref{alg:gMBB} runs much faster.


\begin{algorithm}[t]
\ForEach{$H\in \mathcal{H}$}{

        $H' \leftarrow$ reduce $H$ to $(|A^{*}|+1)$-core\; \label{mv:p}
       let $u$ denote the vertex used to generate $H'$\;
        \tcp{suppose $u\in L$}
       $(A,B)$ $\leftarrow$ $(\{u\},\emptyset)$,  $L(H')$ $\leftarrow$ $L(H')\setminus\{u\}$ \; \label{mv:r1}
       $\small (A', B')$ $\small \leftarrow$ \textsf{\small denseMBB}$((A,B),(L(H'),R(H')))$\; \label{mv:r2}
       \If{$|A'|> |A^{*}|$}{
            $(A^{*}, B^{*})$ $\leftarrow$ $(A',B')$\;
    	}
}

\textbf{return} $(A^{*}, B^{*})$\;
\caption{\textsf{verifyMBB}$((A^{*},B^{*}), \mathcal{H})$}\label{alg:mv}
\end{algorithm}

\section{Experimental studies}\label{sec:exp}
We conduct extensive experiments to verify the effectiveness and efficiency of the proposed techniques and algorithms.

\noindent\textbf{Implemented algorithms}.
We first introduce implemented algorithms evaluated throughout the experimental studies.

\noindent\underline{\textit{Algorithms for dense bipartite graphs}}. We implement Algorithm~\ref{alg:rbb}, denoted as denseMBB. We also implement the state-of-the-art \textsf{MBB} algorithm extBBCL~\cite{ZHOU2018834} as a baseline for comparison.

\noindent\underline{\textit{Algorithms for sparse bipartite graphs}}. We implement our proposed Algorithm~\ref{alg:gMBB} including all the proposed techniques, denoted as hbvMBB.
Besides extBBCL, we use the combinations of existing heuristic \textsf{MBB} algorithms, MBE algorithms and our proposed framework for sparse bipartite graphs to build a number of non-trivial baselines. Before showing these baselines, we discuss the implemented state-of-the-art heuristic \textsf{MBB} and MBE algorithms first.

\textit{Existing heuristic \textsf{MBB} algorithms}. We consider the state-of-the-art heuristic \textsf{MBB} algorithms \textit{POLS}~\cite{wang2018new} and \textit{SBMNAS}~\cite{LI2020104922} for designing non-trivial baselines.The parameter settings are the same as the original papers.

\textit{Existing MBE algorithms}. We adapt existing MBE algorithms by removing maximality and duplication checking. Instead, our proposed upper bound and  the \textsf{MBB} found up to the time will be used to terminate unpromising branches to avoid the costly computations caused by maximality and duplication checking.
The implemented MBE algorithms include \textit{iMBEA}~\cite{zhang2014finding} and \textit{FMBE}~\cite{8990406}.

\textit{Adapted non-trivial baselines}. We use \textit{POLS} or \textit{SBMNAS} to replace the heuristic algorithm used in step 1 of Algorithm~\ref{alg:gMBB} and use the adapted \textit{iMBEA} or \textit{FMBE} to replace our proposed step 2 and step 3 of Algorithm~\ref{alg:gMBB}. As such, algorithms adp1 to adp4 are derived as our baselines shown in Table~\ref{tab:algs}. Please note that the heuristic algorithms that we used are for pruning purpose only, which are discussed in Section~\ref{sec:handr}.

\textit{Variants of our algorithms}. We also implement different variants of our proposed Algorithm~\ref{alg:gMBB}. They are for breaking down evaluations of the proposed techniques, denoted as bd1 to bd5. Their detailed configurations are shown in Table~\ref{tab:algs}.

\noindent\underline{\textit{Other algorithms}}. We also implement the degeneracy algorithm denoted as degOrder to compare with our proposed bidegeneracy algorithm.

\begin{table}[t]
\caption{Implemented algorithms}\label{tab:algs}
\vspace{-10pt}
\small
\begin{adjustwidth}{0cm}{}
\begin{tabular}{|c|l|}
  \hline
  \multicolumn{1}{|c|}{\bfseries Algorithm} & \multicolumn{1}{|c|}{\bfseries Configuration} \\ 
  \hline
  denseMBB	& Algorithm~\ref{alg:rbb}\\
  \hline
  extBBCL & Algorithm proposed in \cite{ZHOU2018834}\\
  \hline
   hbvMBB	&	Algorithm~\ref{alg:gMBB} with all the proposed techniques \\
  \hline
  adp1 & POLS, core based upper bound, FMBE \\
  \hline
  adp2 & POLS, core based upper bound, iMBEA \\
  \hline
  adp3 & SBMNAS, core based upper bound, FMBE \\
  \hline
  adp4 & SBMNAS, core based upper bound, iMBEA \\
  \hline
  bd1  & hbvMBB without step 1\\
  \hline
  bd2  & hbvMBB without core and bicore based optimizations \\
  \hline
  bd3 & hbvMBB without branching technique\\
  \hline
  bd4 & hbvMBB with degree order \\
  \hline
  bd5 & hbvMBB with degeneracy order \\
  \hline
   degOrder & Algorithm for degeneracy order\\
  \hline

\end{tabular}
\end{adjustwidth}
\vspace{-15pt}
\end{table}

\noindent\textbf{Measures}. We measure the running time of the algorithms. The reported running time is the total CPU time (in \textit{seconds}), excluding the I/O cost of loading graph and indices from disk to main memory, and a timeout of $4$ hours is set, denoted as `-'.
All algorithms are implemented in C++. All the experiments are conducted on a PC with CPU of AMD 3900x (12 cores, 24 threads), memory of 64GB DDR4 3600HZ, and Windows 10 (build 1803). All the experiments are conducted no less than 100 times if the running time is less than 1 hour (10 times otherwise) and the average results are reported.

\subsection{Evaluations on Dense Graphs}
\begin{table*}[t]
\small
\centering
\caption{Efficiency for dense bipartite graphs}\label{tab:denseEffi}
\vspace{-5pt}
\begin{tabular}{|c|c|c|c|c|c|c|c|c|c|c|}
 \hline
 \multirow{2}{*}{}& \multicolumn{2}{|c|}{$128 \times 128$} & \multicolumn{2}{|c|}{$256 \times 256$} & \multicolumn{2}{|c|}{$512 \times 512$} & \multicolumn{2}{|c|}{$1024 \times 1024$} & \multicolumn{2}{|c|}{$2048 \times 2048$}  \\
 \cline{2-11}
  & extBBCl & denseMBB                   & extBBCl & denseMBB                 & extBBCl & denseMBB                 & extBBCl & denseMBB                   & extBBCl & denseMBB            \\
 \hline
  \hline
 $70\%$&3.42     &0.869   &682  &3.47    &-      &13.89    &-      &55.58 &-   &232.34        \\\hline
 $75\%$&5.34     &0.854   &1657 &3.45    &-      &13.82    &-      &54.98 &-   &221.14                    \\\hline
 $80\%$&8.32     &0.859   &4025 &3.43    &-      &13.74    &-      &55.21 &-   &219.54           \\\hline
 $85\%$&12.96     &0.854  &9775 & 3.42   &-      &13.67    &-      &54.62 &-   &218.65           \\ \hline
 $90\%$ & 20.2    &0.849  &-  &3.39    &-      &13.59    &-      &54.31 &-   &217.56           \\\hline
 $95\%$ &31.49    &0.845  &-  &3.38    &-      &13.52    &-      &54.12 &-   &211.32           \\\hline
\end{tabular}
\end{table*}

\begin{table*}[t]
\small
  \caption{Efficiency for sparse bipartite graphs}\label{tab:effi}
  \vspace{-5pt}
 \begin{adjustwidth}{-1cm}{}
  \begin{tabular}{ |r|r|l|r|l|r|l|r|l|l|r|}

  \hline
  Dataset   &$|L|$  &$|R|$  &Density$\times 10^{-4}$    &Optimum  &adp1 &adp2 &adp3 &adp4  &extBBCl    &hbvMBB\\
  \hline
   \hline
  unicodelang   &254    &614    &8.0    &$4$  & 0.018 &0.031 &0.0089 &0.0097  &0.0098   &\textbf{0.0071}, S1  \\
  \hline
  moreno-crime-crime &829   &551    &3.2    &2  &0.048 &0.12 &0.038 &0.046  &0.043  &\textbf{0.01}, S1 \\
  \hline
  opsahl-ucforum &899 &522 &71.855 &5 &7.96 & 33.12&5.48 &7.12 &6.43&\textbf{3.21}, S2\\
  \hline
  escorts &10,106 &6624 &0.756 &6 &9.66&41.36 &4.61 &5.24&5.43&\textbf{2.29}, S2\\
  \hline
  jester    &173,421 &100 &563.376 &100 & 978.36 &1784.26 &104.46 &134.65 &1042.57 &\textbf{17.1}, S3\\
  \hline
  pics-ut &17,122 &82,035 &1.637 &30 & 1045.56 &- &746.51 &849.23 &- &\textbf{33.54}, S3 \\
  \hline
  youtube-groupmemberships &94,238 &30,087 &0.103 &12 & 148 &64.5 &8.7& 9.78& 19.63&\textbf{1.25}, S2\\
    \hline
  dbpedia-writer &89,356 &46,213 &0.035 &6 &0.16  &0.26 &0.15 &0.19 & 0.28&\textbf{0.09}, S1\\
    \hline
  dbpedia-starring &76,099 &81,085 &0.046 &6 &2.98  &3.78 &2.12 &2.67 &3.21&\textbf{0.78}, S1 \\
    \hline
  github &56,519 &120,867 &0.064 &12 &112.79&154.13 &98.76 &104.56 & 110.24&\textbf{16.78}, S3\\
    \hline
  dbpedia-recordlabel &168,337 &18,421 &0.075 &6 & 15.13&21.26 &10.24 &13.25 &18.64&\textbf{5.47}, S3\\
    \hline
  dbpedia-producer &48,833 &138,844 &0.031 &6 & 16.29&19.56 &15.46 &15.98 & 17.68&\textbf{8.21}, S3\\
    \hline
  dbpedia-location &172,091 &53,407 &0.032 &5 &0.31 & 0.71& 0.29&0.32 & 0.3& \textbf{0.12}, S1\\
    \hline
  dbpedia-occupation &127,577 &101,730 &0.019 &6 &0.96  &1.21 &0.88 &0.93 &1.1&\textbf{0.12}, S1\\
    \hline
  dbpedia-genre &258,934 &7783 &0.230 &7 &4.52&5.32 &3.81 &4.15 & 4.67&\textbf{0.21}, S1\\
    \hline
  discogs-lgenre &270,771 &15 &1021.2 &15 &0.84&0.99 &0.79 &0.81 & 0.91&\textbf{0.076}, S1\\
    \hline
  bookcrossing-full-rating &105,278 &340,523 &0.032 &13 &351.6  &412.25 &211.47 &311.25 &384.56&\textbf{21.18}, S3\\
    \hline
  flickr-groupmemberships &395,979 &103,631 &0.208 &47 &-& -&784.65 &- &-&\textbf{36.87}, S3\\
    \hline
  actor-movie &127,823 &383,640 &0.030 &8 &1373.67 &2135.62 &885.24 &191.74 &1353.15&\textbf{57.32}, S3\\
    \hline
  stackexchange–stackoverflow &545,196 &96,680 &0.025 &9 &211.54&278.46 &115.46 &189.21 &234.48&\textbf{13.28}, S3\\
    \hline
  bibsonomy-2ui &5,794 &767,447 &0.575 &8 &15.14&22.21 &9.68 &10.12 & 12.47&\textbf{0.87}, S2\\
    \hline
  dbpedia-team &901,166 &34,461 &0.044 &6 &245.47&278.35 &125.67 &198.43 &221.07&\textbf{4.59}, S3\\
    \hline
  reuters &781,265 &283,911 &0.273 &51 &-&- &236.78 &327.24 &-&\textbf{12.35}, S3\\
    \hline
  discogs-style &1,617,943 &383 &38.868 &42 &- &- &587.56 & &-&\textbf{36.47}, S3\\
    \hline
  gottron-trec &556,077 &1,173,225 &0.128 &101 &- &- &156.35 &214.26 &-&\textbf{22.67}, S3\\
    \hline
  edit-frwiktionary &5017 &1,907,247 &0.773 &19 &146.75&352.25 &98.45 &113.59 &125.67&\textbf{6.86}, S3\\
    \hline
  discogs-affiliation &1,754,823 &270,771 &0.030 &26 &1478.56&2574.71 &982.12 &1157.56&1568.43&\textbf{78.53}, S3\\
    \hline
  wiki-en-cat &1,853,493 &182,947 &0.011 &14 &23.23&32.45& 17.56&21.56&24.56&\textbf{1.76}, S2\\
    \hline
  edit-dewiki &425,842 &3,195,148 &0.042 &49 &-&-&89.56 &-&-&\textbf{11.25}, S3\\
    \hline
  dblp-author &1,425,813 &4,000 &0.002  &10 &412.25&578.56 &175.65 &196.48&384.24&\textbf{1.36}, S2\\
  \hline
  \end{tabular}
  \end{adjustwidth}
\end{table*}

\noindent\textbf{Datasets}. We generate dense bipartite graphs for simulating real application scenarios by using random bipartite graph generation algorithm similar to \cite{tahoori2006application}. The range of the edge density ($\frac{|E|}{|L|\cdot|R|}$) in our evaluation is from $0.7$ to $0.95$.

For each edge density and a given size, $100$ instances of bipartite graphs are generated and the average running time is reported for each density. The largest synthetic bipartite graph has $2048$ vertices in each side.
Please note that the largest dense bipartite graph used to evaluate the exact \textsf{MBB} algorithm in~\cite{ZHOU2018834} contains $50$ vertices only.

Table~\ref{tab:denseEffi} shows the running time. We only compare with extBBCL since it is the only exact algorithm that can finish within $4$ hours for some of the tested datasets.
The results clearly demonstrate that denseMBB is able to efficiently handle dense bipartite graphs.
The results also show that denseMBB runs near quadratic time as data becoming dense.
Furthermore, the scalability of denseMBB is also much better than that of extBBCL.
denseMBB can find an \textsf{MBB} within $4$ minutes for bipartite graphs that have $2048$ vertices in each side.

\begin{table*}[t]
\small
\centering
  \caption{Efficiency of our techniques on tough datasets}\label{tab:palgs}
  \begin{adjustwidth}{0.3cm}{}
  \centering
  \begin{tabular}{ |r|r|l|r|l|r|l|r|l|r|}
  \hline
  Dataset   &hMBB  & degOrder  &bdegOrder    &bd1    &bd2 &bd3 &bd4 &bd5 &hbvMBB\\
  \hline
  \hline
  jester  &1.37	&1.19	&2.91	&107.86	&131.82	&337.26		&133.54		&85.61&  \textbf{17.10} \\
  \hline
  pics-ut  &2.01	&2.35	&6.37	&157.64	&275.03	&472.91		&144.22		&50.31&  \textbf{33.54} \\
  \hline
  github  &1.01	&1.51	&2.52	&119.14		&159.41	&204.72		&130.88		&46.98& \textbf{16.78} \\
    \hline
  bookcrossing-full-rating  &1.69	&1.91	&4.02	&127.08	&120.73		&436.31		&135.55	&29.65& \textbf{21.18} \\
    \hline
  flickr-groupmemberships  &1.85&2.96&5.18&273.58&303.15&561.94&184.85&155.27& \textbf{36.87} \\
    \hline
  actor-movie  &2.62&3.66&6.80&287.76&355.78&805.73&340.08&130.80& \textbf{57.32} \\
    \hline
  stackexchange–stackoverflow  &0.93&0.92&1.59&98.27&66.42&224.43&127.49&29.22& \textbf{13.28} \\
    \hline
  reuters  &0.86&0.86&1.73&58.05&117.33&188.96&56.81&30.88& \textbf{12.35} \\
    \hline
  discogs-style  &2.19&3.28&4.38&269.84&320.94&590.81&251.64&138.59& \textbf{36.47} \\
    \hline
  gottron-trec  &1.13&1.81&3.40&154.156&183.63&326.45&129.22&124.69& \textbf{22.67} \\
    \hline
  discogs-affiliation  &4.71&6.28&11.78&581.12&416.21&1625.57&746.03&502.59& \textbf{78.53} \\
    \hline
  edit-dewiki &0.79&0.95&1.83&64.13&78.75&131.63&69.75&67.52& \textbf{11.25} \\
    \hline
  \end{tabular}
  \end{adjustwidth}
\end{table*}

\subsection{Evaluations on Sparse Graphs}
\noindent\textbf{Datasets}. We use real datasets from Koblenz Network Collection (KONECT).
30 instances of the datasets are used to evaluate the algorithms discussed above.
These 30 instances were also used in~\cite{ZHOU2018834}.

We demonstrate the running time of adp1 to adp4, extBBCl and hbvMBB in Table \ref{tab:effi} from 6th to 11th column respectively.
Noticeably, hbvMBB outperforms all the other algorithms for all datasets.
For large datasets, such as actor-movie, hbvMBB runs over 200 times faster than extBBCl.
In average, hbvMBB runs several orders faster than all the other algorithms consistently.
For most of the datasets, the running time of adp3 is the runner-up.
This justifies the power of our proposed search framework.
Please notice that adp3 uses the best reported heuristic algorithm for \textsf{MBB} and the best reported MBE algorithm for maximal biclique.
We shall also highlight that our proposed algorithm can finish within $2$ minutes for all the datasets, whereas extBBCl cannot finish within $4$ hours for $6$ of the datasets.
Although adp3 is the runner-up for most of the datasets, for datasets such as discogs-affilliation and pics-ut, it still needs up to 16 minutes to finish.
This further demonstrates the superiority of our proposed algorithm.

\begin{figure}
	\includegraphics[width=8.3cm]{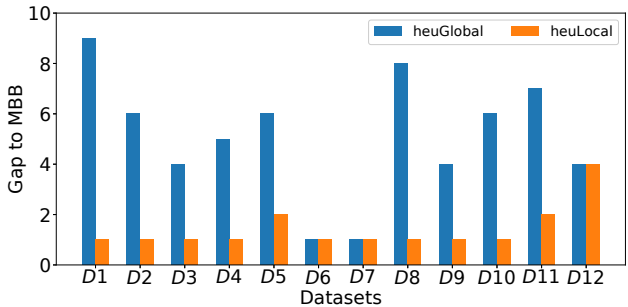}
	\caption{Effectiveness of heuristics}\label{fig:effHeu}
\vspace{-10pt}
\end{figure}

\subsection{Breaking down Evaluations}
In this section, we show the performance of different combinations of our proposed techniques.

\noindent\textbf{Power of heuristics, reduction and early termination}. In Table~\ref{tab:effi} column 11, we show which step hbvMBB terminates.
S1, S2 and S3 mean steps $1$ to $3$ respectively.
For $14$ out of the $30$ datasets, our proposed algorithm, hbvMBB, can terminate at S2.
This is because of two reasons.
Firstly, our proposed heuristics can result in globally maximum balanced biclique.
Secondly, in step $2$, the original graph has been split into vertex centred subgraphs and upper bounds for these subgraphs are significantly tighter.
As such, our proposed early termination conditions have high chance to be satisfied.
This justifies the importance of separating the heuristic from exhaustive search, i.e., it greatly speeds up the search practically and allows an \textsf{MBB} to be found extremely fast in real datasets.
Interestingly, for $8$ datasets, our algorithm terminates at step $1$, which means our algorithm can solve the \textsf{MBB} problem in near linear time for these datasets.

From the above cases we can see that, some datasets are easy to process, which makes them less effective to evaluate our proposed techniques comprehensively.
Therefore, we focus on datasets that $\textsf{hbvMBB}$ cannot finish within $10$ seconds later.

\noindent\textbf{Effectiveness of using different search orders}.
We demonstrate how different search orders, i.e., degree based order, degeneracy order and bidegeneracy order, affect the search performance by reporting the running time of variations of hbvMBB using the three orders.
The results for using degree based order and degeneracy order are shown in Table~\ref{tab:palgs} in columns $8$ and $9$ accordingly.
As we can see, the running time of bd4 and bd5 is slower than that of hbvMBB (using bidegeneracy order) up to 6 times.
Two major reasons cause such dramatic differences.
Firstly, bidegeneracy can make the size of each subgraph that needs to perform exhaustive search smaller than the other two orders.
Secondly, vertex centred subgraph induced by bidegeneracy order can lead to much smaller upper bound compared with the other two orders, i.e., the upper bounds are tighter, which results in better pruning effectiveness.
In addition, the result that bd5 outperforms bd4 confirms that degeneracy order is better than degree order.

\noindent\textbf{Overhead v.s. benefit for core and bicore}.
We report the overhead of computing core (degOrder) and bicore (bdegOrder) for each dataset in Table~\ref{tab:palgs} and the results are shown in columns $3$ and $4$ respectively.
As we can see, the running time of degOrder is trivial and that of bdegOrder is a bit slower for all the datasets.
Note that, during the search, degOrder and bdegOrder are performed on much smaller pieces of data. 
The running time of bd2 (without using any core or bicore based optimizations), is shown in column 6 of Table~\ref{tab:palgs}. 
Compared with bd2, hbvMBB is several times faster than bd2, which justifies that using core and bicore based optimizations can bring dramatic benefits.

\begin{figure}
\vspace{2pt}
	\includegraphics[width=8.3cm]{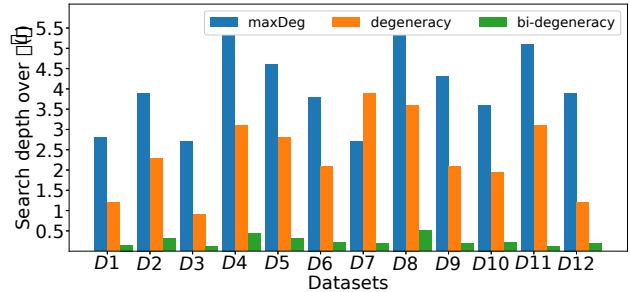}
	\caption{Evaluation on search depth}\label{fig:depth}
\vspace{0pt}
\end{figure}
\noindent\textbf{Overhead v.s. benefit for heuristics}.
We report the running time of heuristic algorithm (hMBB) and the running time of hbvMBB without using hMBB (bd1) on different datasets in columns $2$ and $5$ of Table~\ref{tab:palgs}.
As expected, the running time of hMBB is close to that of degOrder since it is dominated by degOrder.
bd1 takes considerably more time to find an \textsf{MBB} compared to hbvMBB.
This is because hMBB can not only find a large balanced biclique, but also use the found balanced biclique to prune the graph as much as possible.  
From the above analysis, the benefit of using hMBB is significant.

\noindent\textbf{Effectiveness of heuristics}. We report the size gap between the maximum balanced biclique found by our proposed heuristic algorithms and the optimum maximum balanced biclique.
In fact, we have two heuristic parts, i.e., hMBB and the heuristic used in Algorithm~\ref{alg:bridge}, denoted as heuGlobal and heuLocal respectively.
The results are demonstrated in Figure~\ref{fig:effHeu}.
D1 to D12 denote the datasets in Table~\ref{tab:palgs} in top-down order.
As shown, with heuLocal, $9$ out of $12$ datasets can find the global maximum balanced biclique.
This demonstrates the heuristic in Step $2$ can significantly improve the quality of the candidate maximum balanced biclique, which in turn reduces the cost for Step $3$.

\noindent\textbf{Evaluation on search depth}.
We report the average search depth for \textsf{hbvMBB} using different search orders discussed in Lemmas \ref{le:degO}, \ref{le:degenO}, and \ref{le:bdO} denoted as maxDeg, degeneracy and bidegeneracy.
The results are shown in Figure~\ref{fig:depth}.
We use $\ddot{\delta}(\cdot)$ of each of the datasets as a reference and report the ratio of average search depth over $\ddot{\delta}(\cdot)$ for each order.
Overall speaking, the average depth of bidegeneracy is significantly less than those of the other two.
This justifies the size bounding effectiveness of our proposed vertex centred graph.
Noticeably, for all datasets, the average search depth of bidegeneracy over $\ddot{\delta}(\cdot)$ is significantly less than $\ddot{\delta}(\cdot)$, i.e., only 0.12 for D3 and D11.
This justifies the reduction and branching techniques that we propose, and explains why \textsf{hbvMBB} is significantly faster than the other algorithms.

\begin{figure}
	\includegraphics[width=8.3cm]{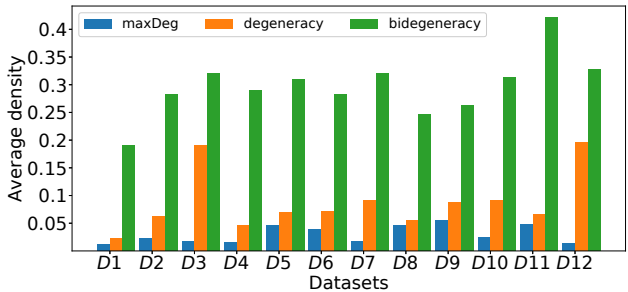}
\vspace{-10pt}
	\caption{Evaluation for vertex centered subgraphs}\label{fig:density}
\vspace{-10pt}
\end{figure}

\noindent\textbf{Evaluation on density of vertex centered subgraphs}.
We report the average density of vertex centered subgraphs generated by different orders denoted as maxDeg, degeneracy and bidegeneracy for each dataset.
The results are shown in Figure~\ref{fig:density}.
Firstly, bidegeneracy has much higher effectiveness to generate high density vertex centered subgraphs. For all datasets, the average density of subgraphs generated using bidegeneracy is an order higher than maxDeg and degeneracy.
For dataset such as D11, the average density of vertex centered subgraph is quite high, i.e., close to a biclique.
This indicates that finding an MBB in such bipartite graph is hard if using the existing techniques that do not optimize for dense subgraphs.

\section{Related works}\label{sec:rw}
Other than the work solving the same problem as we discussed in Section~\ref{sec:sota}, works related to \textsf{MBB} are discussed according to categories below.

\noindent\textbf{Heuristic algorithms for the \textsf{MBB} problem}.
The early works~\cite{al2007defect,tahoori2006application} capture the equivalence of the \textsf{MBB} problem and the maximum independent set problem and use algorithms for independent set problem to solve \textsf{MBB} problem.
In~\cite{ZHOU201986}, an add-swap-drop heuristic algorithm adapted from heuristic maximum clique algorithm~\cite{WU2015693} is proposed. 
In~\cite{wang2018new}, a local search framework based on pair operations is proposed, which swaps pairs of vertices.
Recently, a more general swap based algorithm is proposed~\cite{LI2020104922}, in which multiple vertices can be added, swapped, or dropped at the same time for each of the operations.
None of the above algorithms can guarantee to find the exact result.

\noindent\textbf{Maximum vertex biclique search}.
The problem, finding an maximum vertex biclique (\textsf{MVB}), is polynomially solvable.
Two different techniques are proposed.
In~\cite{10.1006/jagm.2001.1199}, the authors formulate finding an \textsf{MVB} as an instance of integer linear programming (ILP) problem.
In~\cite{denes1931grafok}, the authors discover that finding minimum vertex cover (\textsf{MVC}) in a bipartite graph can be reduced to finding maximum matching in the bipartite graph, where finding maximum matching can be reduced to finding maximum flow in a constructed flow network (based on the bipartite graph).
As such, a polynomial algorithm of finding an \textsf{MVB} in a bipartite graph can be derived since it is equivalent to finding an \textsf{MVC} in the bipartite complementary graph of the bipartite graph.
In fact, maximum matching for bipartite graph can be solved with time complexity of $\mathcal{O}(\sqrt{|L|+|R|}\cdot |E|)$.
Therefore, the \textsf{MVB} problem can be solved in $\mathcal{O}(\sqrt{|L|+|R|}\cdot |E|)$ since  the \textsf{MVB} problem can be reduced to \textsf{MVC} problem.
Different from the \textsf{MVB} problem, the \textsf{MBB} problem cannot be solved in polynomial time due to its NP-hardness. The above techniques cannot be applied to the \textsf{MBB} problem.

\noindent\textbf{Maximum edge biclique search}. The problem of finding maximum edge biclique (\textsf{MEB}) is NP-hard.
Interestingly, very few works study finding an exact \textsf{MEB}.
In~\cite{10.1006/jagm.2001.1199, 8588229}, ILP formulations of \textsf{MEB} are proposed, which can find an \textsf{MEB} with ILP solver.
Since the \textsf{MEB} problem is NP-hard to approximate within a factor of $n^{1-\varepsilon}$, algorithms that can find an \textsf{MEB} with high probability is proposed~\cite{inproceedingsmeb}.
Besides, the \textsf{MEB} problems for special instances of bipartite graphs, i.e., convex bipartite graphs, bipartite permutation graphs and tree convex bipartite graphs, are studied in \cite{citekeymeb} \cite{10.1007/978-3-030-39219-2_10} and \cite{10.1007/978-3-319-59605-1_5} respectively.
Recently, a novel exact \textsf{MEB} algorithm is proposed which can deal with large bipartite graphs.
These techniques cannot be used to discover maximum balanced biclique since there is no balance constraint for the \textsf{MEB} problem.


\noindent\textbf{Maximal biclique enumeration}.
In fact, all variants of maximum balanced biclique problem can be reduced to maximal biclique enumeration.
Besides efficient enumeration, maximal biclique enumeration also focuses on maximality checking and duplication checking.
The state-of-the-art algorithm for maximal biclique enumeration is FMBE proposed in~\cite{8990406}, which is an improvement on LCM-MBC proposed in~\cite{li2007maximal}.
It beats iMBEA proposed in~\cite{zhang2014finding}.
The key improvement is before enumerating all bicliques involved by a vertex, the search scope is reduced to the 2-hop neighbours of the vertex.
In our experiment, we reduce the \textsf{MBB} problem to maximal biclique enumeration with non-trivial adaption.
We have demonstrated that our proposed algorithms outperform algorithms that build on existing techniques.

\section{Conclusion}\label{sec:con}
In this work, we propose novel algorithms for finding maximum balanced biclique in dense bipartite graphs and large sparse bipartite graphs respectively.
For dense bipartite graphs, our proposed algorithm runs in $\mathcal{O}^{*}( 1.3803^{n})$ and runs much faster after applying our proposed optimisations.
For sparse bipartite graphs, our proposed algorithm has the time complexity of $\mathcal{O}^{*}$$( 1.3803^{\ddot{\delta}})$, using our proposed novel techniques.
Extensive experiments are conducted on both synthetic and real datasets to demonstrate the practical performance of the proposed algorithms and the effectiveness of the proposed techniques.



\bibliographystyle{abbrv}
\bibliography{vldb_main}  

\end{document}